\documentclass[preprint,10pt]{elsarticle}

\usepackage{amssymb}
\usepackage{url}
\usepackage{graphicx}
\usepackage{subcaption}
\usepackage{amsmath}
\usepackage{svg}
\usepackage{graphicx} 
\usepackage{makecell}
\usepackage{float}
\usepackage{courier}

\usepackage{xcolor}
\usepackage{listings}
\definecolor{codebg}{RGB}{245,245,245} 
\journal{Computer Physics Communications}

\begin{document}

\begin{frontmatter}

\title{exa-AMD: An Exascale-Ready Framework for Accelerating the Discovery and Design of Functional Materials}

\author[a,b]{Weiyi Xia}
\author[c]{Maxim Moraru}
\author[c]{Ying Wai Li}
\author[a,b]{Cai-Zhuang Wang\corref{author}}
\cortext[author]{Corresponding Author.\\\textit{E-mail address:} wangcz@ameslab.gov}
\address[a]{Ames National Laboratory, U.S. Department of Energy, Iowa State University, Ames, Iowa 50011, USA}
\address[b]{Department of Physics and Astronomy, Iowa State University, Ames, Iowa 50011, USA}
\address[c]{Computer, Computational, and Statistical Sciences Division, Los Alamos National Laboratory, Los Alamos, NM 87545, USA}

\date{June 2025}

\begin{abstract}
We present exa-AMD, an open-source, high-performance framework designed for accelerated materials discovery on modern supercomputers. exa-AMD overcomes key computational bottlenecks in large-scale structure prediction through task-based parallelization, adaptive load balancing, and optimized data management for CPU and GPU architectures. The framework automates the end-to-end workflow—from generating candidate structures to evaluating formation energies and updating phase diagrams. Its modular design allows users to easily replace or extend components with custom machine learning models, alternative initial structure templates, and future structure generators, enabling flexible integration with emerging AI approaches. We demonstrate strong scaling across high-performance computing platforms and highlight applications to Na–B–C, Ce–Co–B, and Fe–Co–Zr systems, establishing exa-AMD as a robust and exascale-ready tool for accelerating the discovery and design of functional materials. exa-AMD is publicly available on GitHub, with detailed documentation and reproducible test cases to support community engagement and collaborative research. 
\end{abstract}


\end{frontmatter}

{\bf PROGRAM SUMMARY/NEW VERSION PROGRAM SUMMARY}

\begin{small}
\noindent
{\em Program Title:} exa-AMD \\
{\em CPC Library link to program files:} (to be added by Technical Editor) \\
{\em Developer's repository link:} \url{https://github.com/ml-AMD/exa-amd/} \\
{\em Code Ocean capsule:} (to be added by Technical Editor)\\
{\em Licensing provisions:} BSD3 \\
{\em Programming language:} Python \\
{\em Supplementary material:} [if any] \\
{\em Nature of problem (approx. 50-250 words):} The discovery of novel functional materials in multinary chemical systems is hindered by the combinatorial explosion of possible compositions and structures, making exhaustive exploration computationally intractable with traditional methods. High-throughput density functional theory (DFT) screening is limited by its immense resource demands, particularly for predicting thermodynamic stability and functional properties in complex spaces like ternary or quaternary alloys where millions of candidates must be evaluated. Additionally, integrating machine learning (ML) acceleration, workflow automation, and exascale resource management into a unified, reproducible framework remains a significant challenge, often resulting in inefficient or non-portable solutions that fail to guide experimental synthesis effectively. \\

{\em Solution method (approx. 50-250 words):} exa-AMD addresses these challenges through a modular, Python-based workflow that automates structure generation via template substitution, rapid stability screening using machine learning models for formation energy prediction followed by DFT calculations for validation. These are managed by the Parsl library for dynamic task distribution across CPU/GPU clusters. The framework efficiently filters candidates by energy thresholds and structural similarity, computes convex hulls for thermodynamic stability assessment, and supports elastic scaling on high performance computing platforms, enabling the discovery of stable and metastable compounds from user-specified elements within hours to days.  \\

{\em Additional comments including restrictions and unusual features (approx. 50-250 words):} Structural motifs have to be provided by users in the initial prototype set (e.g., from databases such as Materials Project or user-provided templates), potentially missing novel structures not represented therein. However, users can fully customize the input structure pool, ML models, or DFT backends for specific applications. The code requires access to compatible quantum simulation software (e.g., VASP) and HPC schedulers (e.g., SLURM), with performance optimized for GPU acceleration in ML and DFT stages, though it runs on CPUs as well.  \\

\end{small}

\section{Introduction}
The discovery of novel functional materials is one of the major scientific challenges of the twenty-first century. The challenge is particularly acute for multinary systems, where the number of potential compositions and crystal structures grows exponentially with the number of constituent elements, making exhaustive searches with traditional trial-and-error and brute-force approaches practically impractical, due to the combinatorial explosion of atomic configurations and the high computational cost of predicting electronic, magnetic, and thermodynamic properties from first principles. To address this, the field has shifted towards a computational paradigm that leverages high-throughput first-principles calculations to accelerate the discovery cycle, systematically screening hypothetical candidates to identify promising materials for targeted synthesis.

Traditional high-throughput methods, primarily based on density functional theory (DFT), are used to establish large materials databases such as the Materials Project~\cite{MP}, Automatic FLOW for Materials Discovery (AFLOW)~\cite{AFLOW}, and the Open Quantum Materials Database (OQMD)~\cite{OQMD,OQMD2}. These approaches focus on computing key properties such as formation energies, band gaps, and elastic moduli for known or enumerated structures, providing valuable insight into phase stability and guiding experimental efforts~\cite{MP, AFLOW, OQMD, OQMD2, high-throughput-review1, high-throughput-review2}. However, they are severely limited in coverage and efficiency for complex systems, as brute-force DFT evaluations become computationally intractable for millions of candidates, often requiring months of supercomputing time and lacking the flexibility to intelligently prioritize low-energy structures without exhaustive calculations.

A solution to this problem is the use of machine learning (ML) models to rapidly predict formation energies and stability metrics for novel materials and down-select candidates before performing costly DFT validation, thereby accelerating the exploration of significantly larger chemical and structural spaces. ML models such as graph neural networks have shown promises in achieving DFT accuracy at a fraction of the cost, reducing screening time from months to minutes while maintaining predictive reliability~\cite{CGCNN, ML-materials-review}. This data-driven acceleration is essential for tackling scientific problems like designing rare-earth-free magnets or novel battery materials, where multi-objective optimization across stability, magnetism, and synthesizability is required.

In this work, we present \textbf{exa-AMD}, an open-source, modular, and scalable framework for Accelerated Materials Discovery targeted to exascale platforms. exa-AMD unifies data mining, machine learning, advanced workflow automation, and first-principles computation into a holistic framework, enabling rapid prediction of stable and metastable compounds and property optimization in complex, multi-element chemical spaces that are otherwise inaccessible through conventional empirical or brute-force computational methods. It supports ternary and quaternary alloys, and can be extended to higher-order systems. exa-AMD integrates advanced ML models and high-throughput quantum mechanical calculations to enable rapid screening and characterization of novel compounds through dynamical computing resource management. The software streamlines the entire materials discovery process, from structural hypothesis generation to thermodynamic stability and property prediction, thereby significantly reducing the time and resources needed for discovering promising candidates. The code's flexible, modular design supports various workflow needs for researchers to efficiently target new compounds for synthesis and technological applications, as will be demonstrated by two examples in designing rare-earth-free magnets and beyond. We also show benchmarking results to demonstrate its excellent scaling on some of the largest high performance computing (HPC) resources available.

\section{Program Overview}
\begin{figure}[htb]
\centering
\includegraphics[width=1.0\textwidth]{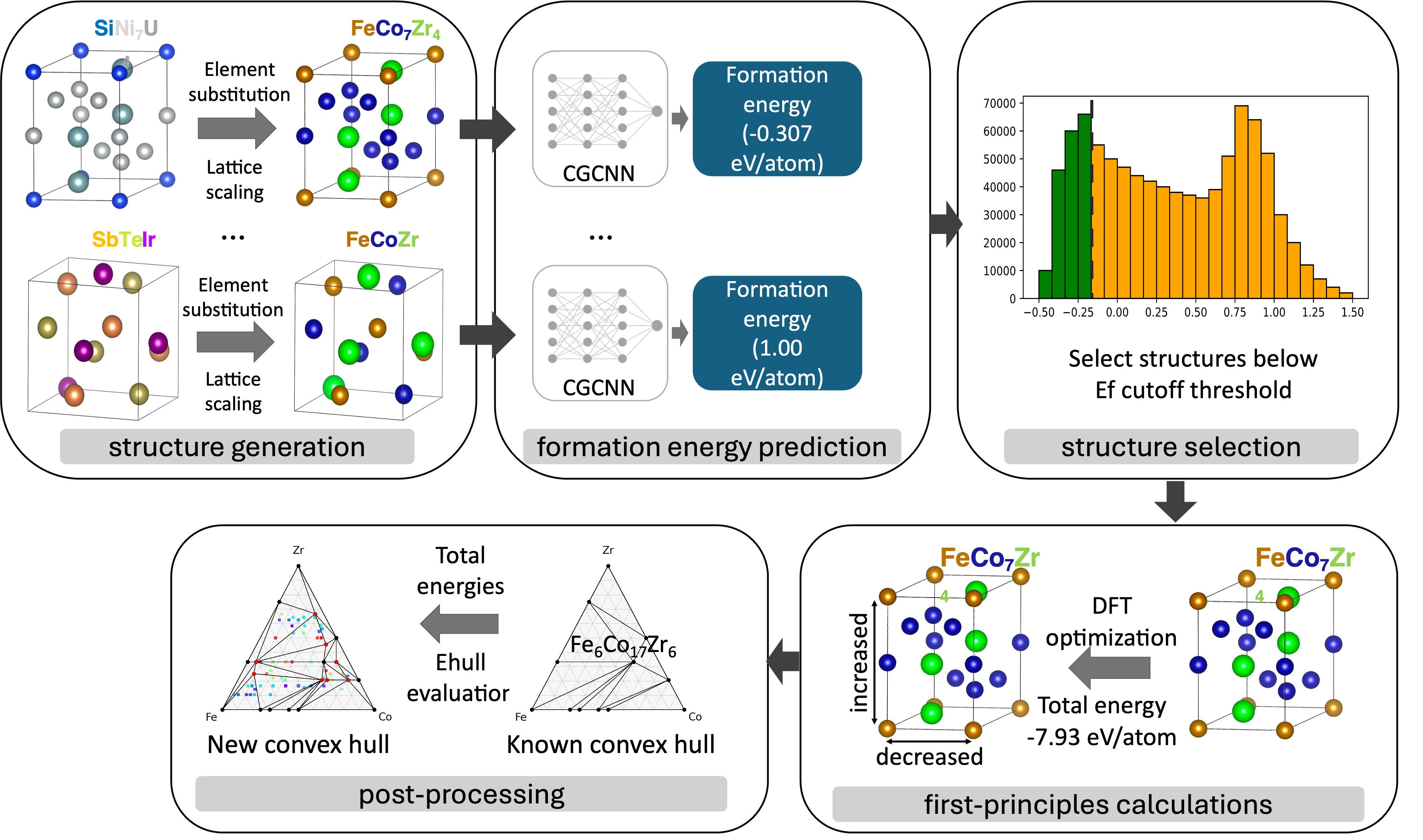}
\caption{Schematic workflow consists of five major steps: (1) crystal structures construction; (2) formation energy prediction using ML models; (3) structure selection; (4) first-principles calculations; (5) post-processing.}
\label{fig:exaAMD-overview}
\end{figure}

The exa-AMD framework is designed as a modular, high-throughput pipeline for accelerated materials discovery, automating the entire process from user-specified chemical elements to the prediction of stable compounds and phase diagrams. Implemented in Python, it leverages ML for rapid stability screening and DFT for precise characterization, while ensuring extensibility, reproducibility, and efficient scaling from laptops to GPU-rich HPC environments. The basic workflow is comprised of five major steps as shown in Fig.~\ref{fig:exaAMD-overview}:
(1) crystal structures construction;
(2) formation energy prediction using ML models;
(3) duplicate removal and structure selection;
(4) first-principles calculations and characterization of physical properties; 
(5) post-processing which assesses thermodynamic stability via convex hull construction, and generates updated phase diagrams.
Each of these stages is encapsulated as a configurable module, allowing users to adapt to new chemical systems, swap ML models, or integrate alternative quantum calculation packages, with workflow orchestration handled by Parsl~\cite{Parsl} for dynamic resource management. In the following, we will describe each step in details.

\subsection{Crystal structures construction}

At the beginning of the framework, an initial pool of crystal structures in the Crystallographic Information File (CIF) format should be provided. They are the ``seeds'' from which the new materials will be generated. These crystal structure files can be obtained from existing databases (Materials Project~\cite{MP}, GNoME~\cite{GNOME}, OQMD~\cite{OQMD,OQMD2}, AFLOW~\cite{AFLOW}, and NovoMag~\cite{Novomag}, ensuring a broad coverage of known and plausible motifs. By default, exa-AMD provides an initial pool of 36553 unique ternary structures, and 5254 unique quaternary structures. Users can also prepare their own prototype crystal structures as their customized initial structure pool of their research interest. 

Hundreds of thousands to nearly one million candidate structures are then generated using two complementary methods, visually represented in Fig.~\ref{fig:structural-generation}:
\begin{figure}[h!tb]
    \centering

    \begin{subfigure}[t]{0.49\textwidth}
        \centering
        \includegraphics[width=\linewidth]{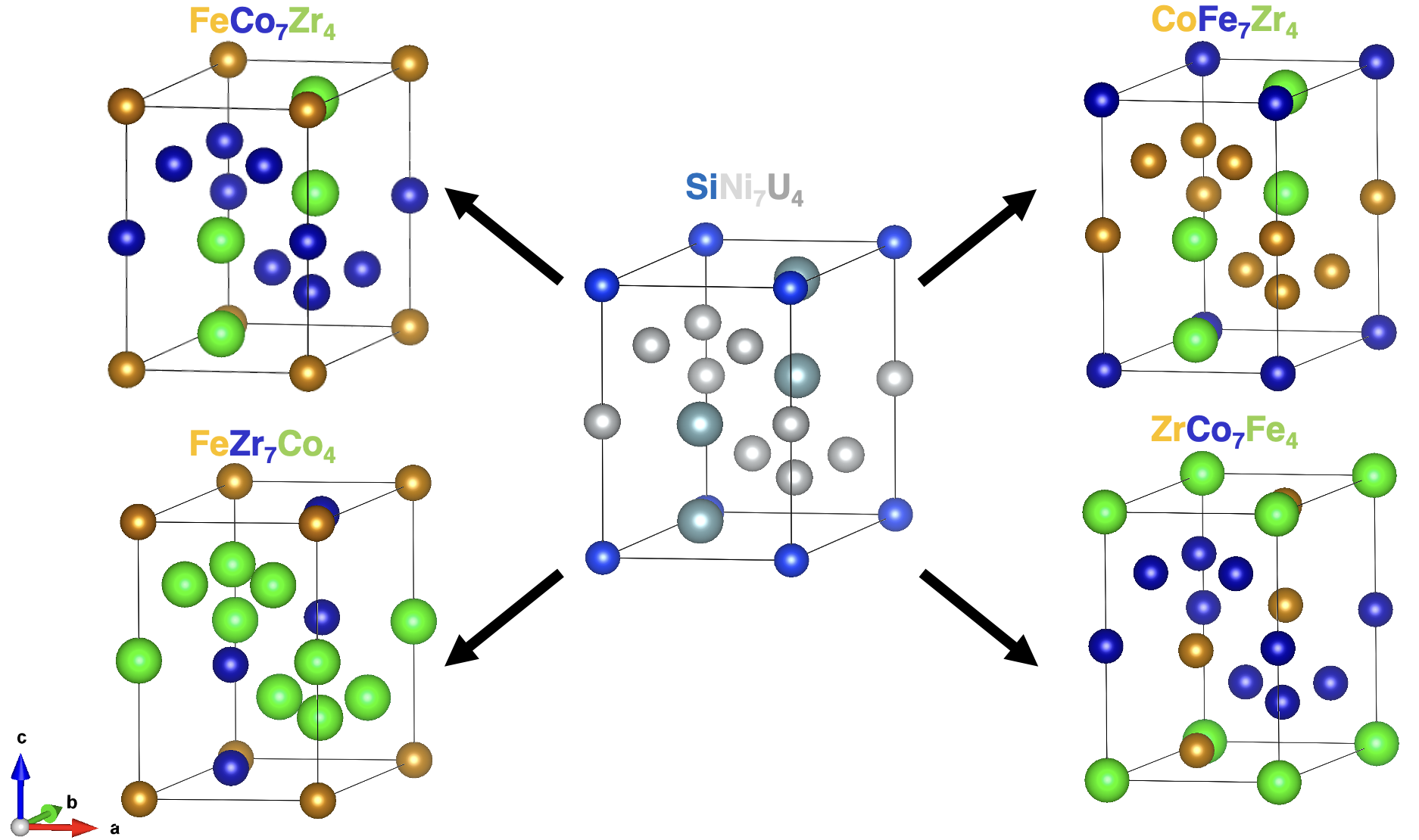} 
        \vspace{5pt}
        \caption{Element Substitution.}
        \label{subfig:element-substitution}
    \end{subfigure}\hfill
    \begin{subfigure}[t]{0.49\textwidth}
        \centering
        \includegraphics[width=\linewidth]{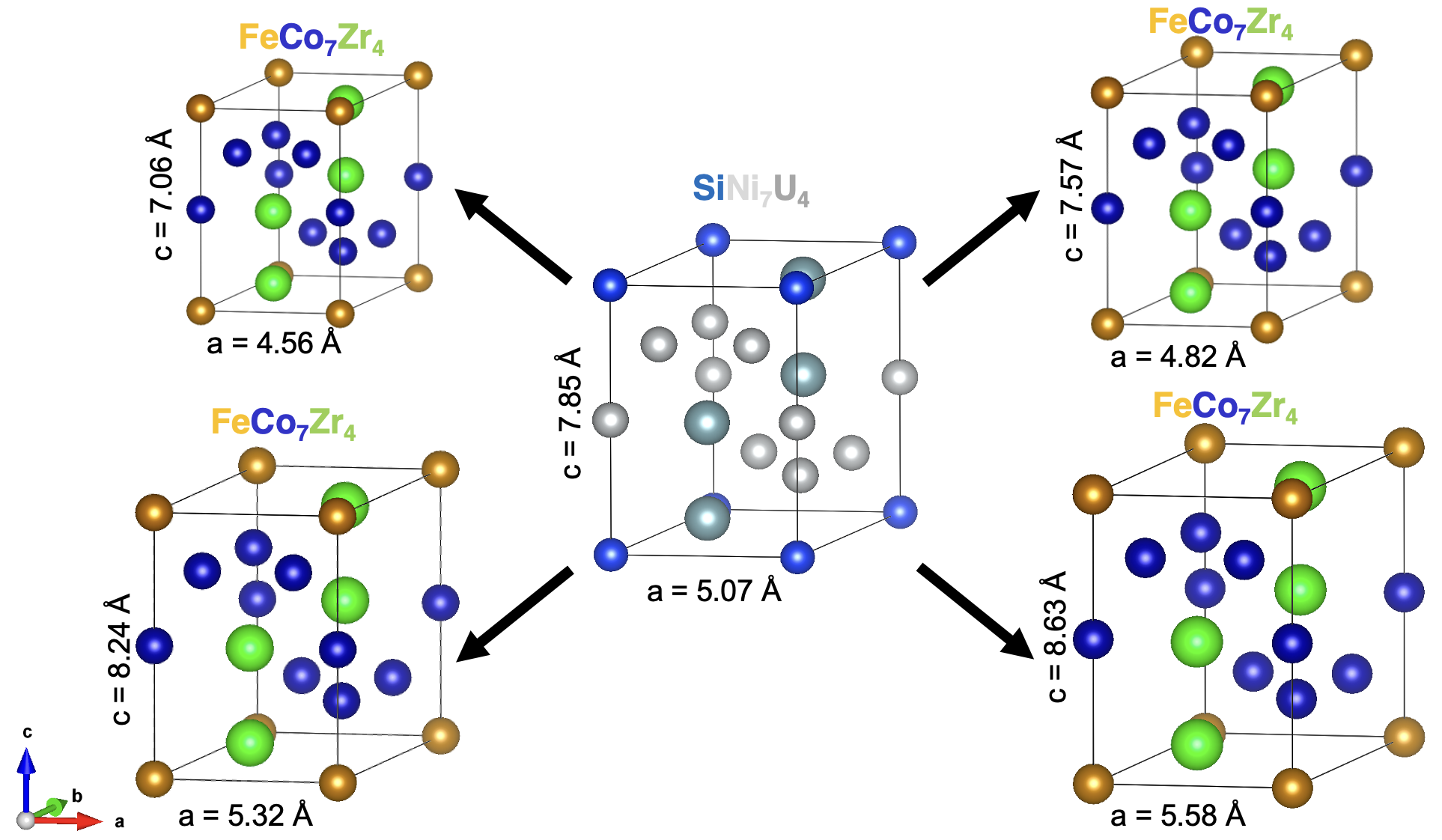} 
        \vspace{5pt}
        \caption{Lattice-Volume Scaling.}
        \label{subfig:lattice-scaling}
    \end{subfigure}

    \vspace{5pt}
   \caption{
    Illustration of the two primary high-throughput structure generation methods. \textbf{(a) Element substitution}: systematic replacement of atomic species at fixed crystallographic sites. \textbf{(b) Lattice-volume scaling}: uniform expansion or contraction of the unit cell volume by applying scaling factors.
    }
    \label{fig:structural-generation}
\end{figure}
\subsubsection*{(i) Systematic elemental substitution}
Systematic elemental substitution with combinatorial atom-type shuffling is used to explore compositional and configurational space (Fig.~\ref{fig:structural-generation}a). In this process, each prototype structure undergoes the systematic replacement of its atomic species with the user-specified target elements, while simultaneously permuting the element assignments across available crystallographic sites~\cite{FeCoZr,CeCoCu,JOSS}. For example, for a ternary template structure A-B-C, and the user specifies target elements X, Y, and Z, the code generates all chemically feasible combinations (e.g., X-Y-Z, X-X-Z, Y-Y-Z, etc.) by substituting each site with each target element. This approach ensures comprehensive coverage of both compositional space and configurational degrees of freedom within each structural motif, while preserving the underlying crystal symmetry and topology of the prototype structure.
\subsubsection*{(ii) Lattice-volume scaling}
Lattice-volume scaling is applied to approximate the equilibrium bond lengths for the newly generated compounds. (Fig.~\ref{fig:structural-generation}b). Since the atomic sizes of the elements in the template structure may differ significantly from those in the target system, the equilibrium volumes of the substituted compounds will also differ substantially. To account for this, scaling factors are applied: typically multiple values ranging from 0.90 to 1.10, which uniformly expand or contract the unit cell volume. For example, a prototype with lattice constant \(a_0\) is scaled to \(0.92a_0\), \(0.96a_0\), \(1.0a_0\), \(1.04a_0\), and \(1.08a_0\), as described in \cite{JOSS, FeCoZr, CeCoCu}.


\subsection{Rapid stability screening using ML models}

To efficiently predict the formation energies and eliminate high-energy candidates, exa-AMD employs an ML model, Crystal Graph Convolutional Neural Network (CGCNN), for high-throughput stability screening~\cite{CGCNN}. CGCNN represents each hypothetical crystal as a graph, encoding atoms as nodes (featuring atomic number and chemical environment) and bonds as edges (including interatomic distances). Each hypothetical structure generated from the previous step is evaluated using the CGCNN model that is trained to predict the formation energy. The initial, universal CGCNN model we utilized in this framework was trained based on 28046 structures from Materials Project, as described in~\cite{CGCNN}. The mean absolute error of this model is typically around 0.1--0.2 eV/atom.

Down selection is then carried out to remove duplicates and structures with high energies. First, candidates are sorted by the predicted formation energy, and basic filters (a formation energy cutoff, e.g., $E_\mathrm{f} < 0$~eV/atom, optional element-fraction limits, and a maximum-atoms limit) are applied. To identify unique crystal structures, we then group the remaining candidates by their reduced composition. Within each group, the structures are processed from lowest to highest formation energy. A structure is kept for subsequent DFT calculation only if it does not match any previously kept structure from that group, as determined by the \small{\texttt{pymatgen.analysis.structure\_matcher.StructureMatcher}} class with the default difference tolerance \cite{pymatgen}. This results in typically 1,000--4,000 unique candidates for first-principles calculations in the next step~\cite{FeCoZr,CeCoCu,JOSS}.

Our previous work shows that the use of ML models drastically reduces the computational time from months (using brute force DFT) to minutes for over $10^6$ structures~\cite{CeCoCu,CeFeX,FeCoC,FeCoP,FeCoSi,FeCoZr,LaCoPb,PNAS,JMCA,Sun2022}. This screening enables the rapid exclusion of high-energy or chemically implausible candidates before any quantum calculation is performed.

After the completion of the first round of the framework (i.e., including DFT calculations and post-processing as discussed in Sections~\ref{DFT} and~\ref{postprocessing}), a system-specific, second-generation CGCNN model is retrained using several hundred to thousands of new DFT-relaxed ground states, reducing the mean absolute errors to as low as 0.03--0.05~eV/atom for the chemical systems of interest~\cite{PNAS,JMCA,FeCoZr,CeCoCu,FeCoC}.
Key hyperparameters for training include 3--6 convolutional layers, batch size of 64--256, 100--200 epochs, stochastic gradient decent optimization, and 80/10/10 data splits for training/validation/testing, respectively. In this work, we choose to use the default setting of these parameters provided by Xie, et al \cite{CGCNN}.

Since the evaluation of each candidate structure is independent, this approach provides a scalable and accurate procedure to select the most promising candidate materials for the costly quantum calculations in the next step. Importantly, our framework allows users to replace or augment the ML model with other ML architectures or retrained models specific to their system of interest, enabling improved prediction accuracy and tailored screening strategies. Such flexibility allows adaptation to a wide range of material classes and accelerates convergence to plausible candidate structures.

\subsection{First-Principles Calculations}
\label{DFT}

In exa-AMD, we use density functional theory as implemented in the Vienna Ab initio Simulation Package (VASP)~\cite{VASPa,VASPb}. By default, structural relaxations and total energy calculations employ the projector augmented-wave (PAW) method~\cite{PAW} and the Perdew-Burke-Ernzerhof (PBE) GGA exchange-correlation functional~\cite{GGA}, with a plane-wave cutoff of 520~eV and Monkhorst-Pack $k$-point meshes at a density of 2$\pi \times 0.025$~\AA$^{-1}$ to ensure convergence especially for metals and intermetallics~\cite{FeCoZr,CeCoCu,FeCoC}. These parameters are chosen based on systematic convergence tests to balance accuracy and computational efficiency: the 520~eV cutoff ensures well-converged total energies for 3d transition metals, while the $k$-point density provides sufficient sampling of the Brillouin zone for accurate electronic structure and forces in metallic systems with moderate unit cell sizes. The lattice parameters and internal atomic positions are relaxed until all forces fall below 0.01~eV/\AA. The electronic band structures are also computed with high accuracy. For magnetic systems, spin polarization is included, with the initial moment set to ferromagnetic configuration. The saturation magnetization ($M_s$, $J_s$) is computed from the total moment and relaxed cell volume. All DFT calculation parameters are fully customizable by users through the VASP INCAR file, which is automatically generated by exa-AMD but can be easily modified to accommodate specific research needs. Users can adjust convergence criteria (e.g., EDIFF, EDIFFG), change the exchange-correlation functional (e.g., HSE06 for hybrid functional calculations), apply Hubbard $U$ corrections for strongly correlated systems (via LDAU flags), modify $k$-point meshes, or set non-collinear magnetization for complex magnetic structures. The framework's modular design ensures that custom INCAR settings are preserved across workflow stages while maintaining full automation of job submission and monitoring.

Formation energies per atom are always referenced to the relaxed energies of the elemental (and, if needed, binary) phases, and convex hull constructions are used to assess thermodynamic stability---a compound is designated as stable, metastable ($E_{\rm hull} < 0.1$~eV/atom), or unstable in this step. For correlated systems, a Hubbard $U$ correction is applied within standard DFT for more accurate valence treatments when necessary. Dynamical stability checks, such as phonon calculations and ab initio molecular dynamics, can be conducted for representative new phases as post-processing steps. By parallelizing DFT jobs across CPU/GPU clusters using Parsl, thousands of structure relaxations with varying size and complexity can be completed within hours to days---enabling practical exploration of otherwise intractable structural and compositional spaces~\cite{JOSS,FeCoZr,CeCoCu,PNAS,JMCA}.
Although this first-principles calculation stage primarily uses VASP as the default DFT engine, the workflow is designed to be agnostic about the choice of DFT software. Depending on user preferences and available computational resources, alternative software packages such as Quantum ESPRESSO~\cite{QEa,QEb} can be used. This modularity not only offers flexibility in computation but also enables benchmarking and method comparison to ensure robust validation of candidate materials.

\subsection{Convex Hull and Thermodynamic Stability}
\label{postprocessing}

The formation energy of a candidate material is defined as
\begin{equation}
  E_\text{form} = E_\text{tot}(\text{compound}) - \sum_i n_i E_\text{ref}(i),
\end{equation}
where $E_\text{tot}$ is the total energy of the candidate, $n_i$ are the atomic fractions, and $E_\text{ref}(i)$ are the reference energies of the constituent elements.

Candidates are compared against all known phases to compute their energy above the convex hull ($E_\mathrm{hull}$). To construct the reference convex hull, we retrieve all thermodynamically stable and metastable structures for the relevant chemical system (e.g., Fe-Co-Zr ternary system) from the Materials Project database~\cite{MP} using their Python Application Programming Interface (API). Specifically, we use the \texttt{MPRester.get\_entries\_in\_chemsys()} method to obtain all \texttt{ComputedStructureEntry} objects containing the target elements, which include elemental phases (e.g., pure Fe, Co, Zr), binary compounds (e.g., Fe-Co, Fe-Zr, Co-Zr), and known ternary phases in the chemical system. Each entry contains the DFT-calculated total energy and crystal structure, from which formation energies are computed relative to the elemental reference states. These formation energies, combined with our newly calculated DFT results for the generated candidate structures, are used to construct the convex hull --- the set of thermodynamically stable phases, where any compound above this hull is metastable or unstable~\cite{convexhull}. Compounds on the hull ($E_\mathrm{hull}=0$ eV/atom) are identified as thermodynamically stable, while those near the hull (e.g., $E_\mathrm{hull} < 0.1$ eV/atom) are considered metastable and potentially synthesizable under non-equilibrium conditions. This rigorous convex hull analysis is essential for predicting which compounds can potentially be synthesized in practice, as confirmed by literature and data-mined studies~\cite{AFLOW,GNOME}.
After all candidate structures complete DFT relaxation and their formation energies are computed, the convex hull is updated by combining the newly calculated structures with the retrieved existing known phases. Newly discovered low-energy structures may shift the convex hull downward, potentially reclassifying previously stable Materials Project phases as metastable, while simultaneously establishing new stable or near-stable phases. The final output represents the most complete and up-to-date phase diagram for the chemical system, incorporating both the Materials Project reference data and all validated compounds from the current discovery campaign.

It is important to note that the scope of the predicted phases and phase diagrams generated by exa-AMD is fundamentally determined by the set of structure prototypes supplied at the beginning of the workflow. These initial prototypes, either provided by default with exa-AMD or customized by the user, define the structural motifs accessible to elemental substitution, combinatorial screening, and all subsequent ML and DFT evaluations. Consequently, all new compounds, low-energy structures, and phase diagram updates produced by the current version of exa-AMD are restricted to the chemical and structural space spanned by these templates. While this approach enables high throughput calculations and chemical flexibility, it does not guarantee exhaustive exploration of all possible structure types. As demonstrated in our previous work on the Fe-Co-C system~\cite{FeCoC}, this limitation can be addressed by integrating adaptive genetic algorithms (AGA)~\cite{AGA1,AGA2} capable of exploring and predicting new structure types that are absent from existing databases or user-supplied templates. In future releases of exa-AMD, we plan to natively incorporate AGA-based structure prediction to overcome the prototype bottleneck and expand the framework’s reach to uncover genuinely novel structural motifs and compounds beyond the current prototype-driven paradigm.

\section{Software Implementation}
exa-AMD is implemented in \texttt{Python} and designed for clarity, portability, and extensibility. It encompasses plug-in scripts for structure handling, database access, machine learning inference, and DFT job submission. The use of the Parsl library~\cite{Parsl} ensures efficient, scalable workflow orchestration involving task parallelism, fault tolerance, and elastic resource management. Users can configure resource allocation, execution platforms, and job scheduling via flexible configuration files, allowing seamless portability from local workstations to large-scale HPC systems. Key features of the software implementation include:
\begin{itemize}
\item \textbf{Modular design:} The framework's modular architecture allows users to customize or replace key components to adapt to diverse research needs. For example, users can provide their own customized initial structure pool as discussed above. Moreover, the machine learning model (default is CGCNN currently), can be replaced by other advanced models with the same interface, such as ALIGNN \cite{ALIGNN}, M3GNet\cite{M3GNET}, MEGNet\cite{MEGNET}, CHGNet\cite{CHGNET}, etc. The first-principles component also supports multiple DFT engines such as VASP, Quantum ESPRESSO, and others.

\item \textbf{Workflow orchestration:} Our framework manages scheduling, distribution, and monitoring of tasks, supporting both CPU and GPU execution with elasticity and fault tolerance to efficiently use heterogeneous computing nodes.
\item \textbf{I/O and database integration:} Structures are sourced from our exa-AMD dataset \cite{exaAMD-dataset}, curated from multiple databases or user-provided datasets, with outputs systematically organized into logs, results, and configuration files to ensure comprehensive data provenance and reproducibility.
\item \textbf{Parallel execution:} High-throughput tasks like ML predictions and DFT calculations are automatically parallelized, with configurable resource allocations (node types and counts) for elastic scaling on supercomputers and clouds.
\item \textbf{User interface:} Flexible command-line tools and configuration files support simple and advanced job submission, including the ability to restart workflows at any stage.
\item \textbf{Documentation and testing:} Comprehensive user guides, API documentation, and automated test suites are provided to facilitate adoption and ensure software reliability.
\end{itemize}

At the heart of exa-AMD’s scalability and automation is the use of Parsl~\cite{Parsl}, a flexible, Python-based parallel programming and workflow library designed to efficiently orchestrate scientific pipelines across heterogeneous computational resources. Parsl enables exa-AMD to decompose each major workflow step---structure generation, ML-based screening, structure similarity filtering, DFT calculations, and property post-processing---into fine-grained, asynchronous tasks (``apps'' in Parsl terminology) that can be dynamically mapped onto available compute resources such as CPUs and GPUs. 

\begin{figure}[H]
  \centering
  \includegraphics[width=\linewidth]{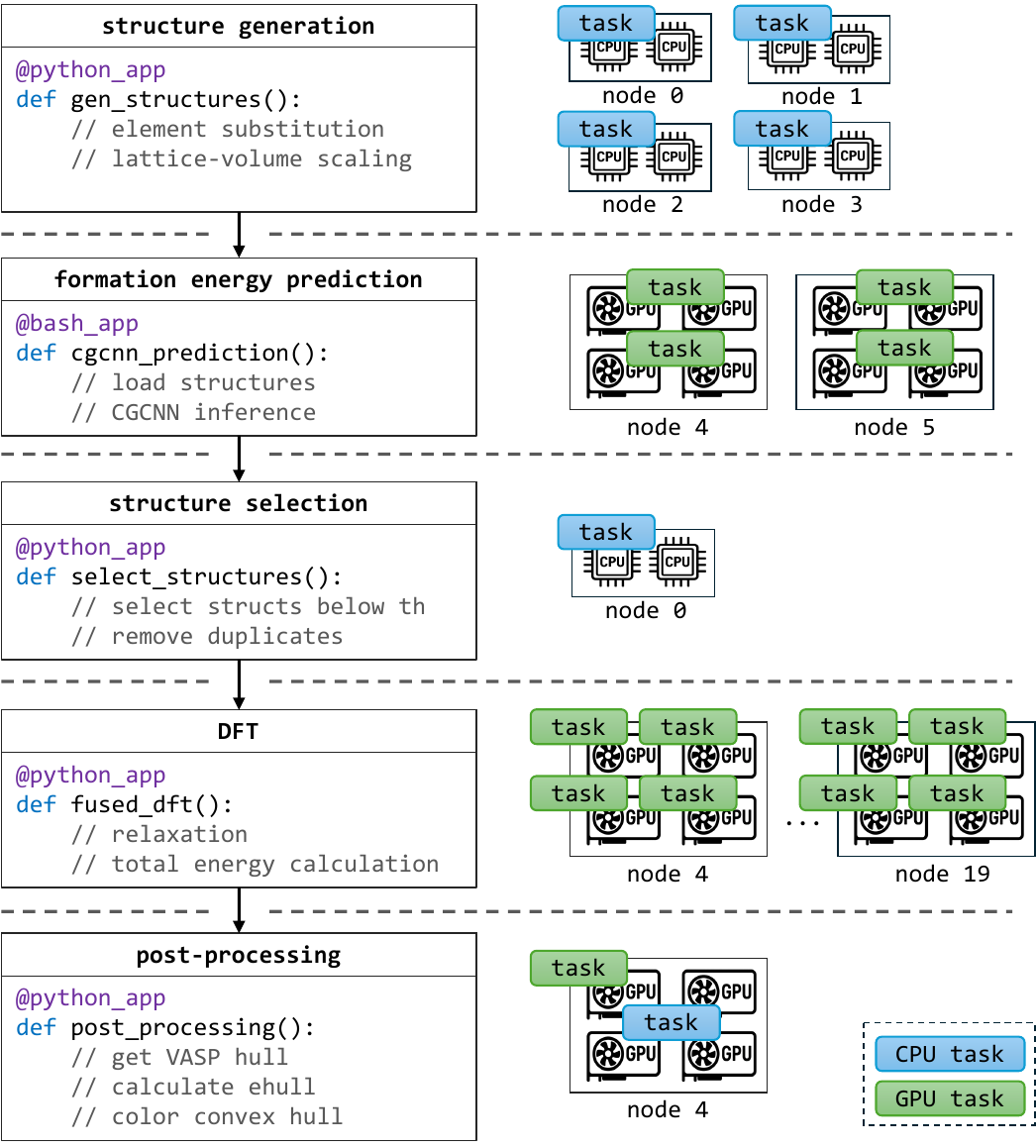}
  \caption{Example mapping of exa-AMD stages to hardware via Parsl executors: CPU nodes (blue) and GPU nodes (green).}
  \label{fig:workflow-hw-mapping}
\end{figure}

Building on this task-based execution model, users provide a Parsl configuration that operates independently of the scientific workflow itself. The configuration defines how each workflow stage is mapped onto computational resources. Fig.~\ref{fig:workflow-hw-mapping} shows an example mapping on a heterogeneous machine where CPU nodes are dual-socket and GPU nodes contain four accelerators each. The figure illustrates several key aspects of exa-AMD's execution model. 

First, at every stage of the workflow, exa-AMD achieves scalability through task-level concurrency across nodes and intra-node parallelism. For example, independent tasks, such as VASP calculations on different candidate structures, are executed concurrently across multiple nodes, while multithreading and GPU acceleration exploit parallelism within each node. 

Second, certain stages (structure generation and formation-energy prediction) are directly parallelized through Parsl. Users can specify how the overall workload is partitioned into independent chunks, which determines the number of concurrent tasks. In the example shown in Fig.~\ref{fig:workflow-hw-mapping}, the workload is divided into four chunks, producing four parallel tasks.

Third, the execution model does not require dedicating one node per task. Multiple lightweight tasks can share the computational resources of a single node. For instance, four independent tasks may run concurrently on a single GPU node. The user determines the allocation strategy according to the size and complexity of the target compound, balancing throughput against resource efficiency. In practice, the DFT stage dominates the overall computational cost (see Section~\ref{example_applications}), which is reflected in the figure by allocating sixteen GPU nodes to this stage.

Fourth, the post-processing stage can reuse the existing GPU allocation to avoid additional wait time in the queue on a cluster. In this stage, exa-AMD performs VASP single-point total-energy calculations on GPUs, followed by short CPU-only tasks such as computing and plotting the convex hull. For such brief operations, reusing the same GPU allocation is generally more efficient than requesting a separate CPU allocation, as it eliminates unnecessary scheduling delays.

In addition, for performance reasons, the number of concurrent tasks being processed at a time, used in the ML prediction stage should be an integer divisor of that used in the structure generation stage, because the initial partitioning of the problem is established at the first stage. In the example of Fig.~\ref{fig:workflow-hw-mapping}, the initial workload is divided into four chunks and thus produces four tasks. Structure generation runs on four nodes with full concurrency, while the ML prediction stage, using only two nodes, executes the same four tasks in two successive waves of two concurrent jobs.


Parsl also enables robust resumability and error handling within exa-AMD. Every task’s state and results are tracked, allowing failed calculations to be retried or workflows to resume from intermediate stages in the event of an interruption.
This highly modular and fault-tolerant approach not only maximizes resource usage and throughput on shared HPC queues but also supports rapid development and reproducibility---key requirements for cutting-edge, community-driven materials discovery frameworks.

By decoupling workflow logic from execution configuration, exa-AMD with Parsl facilitates portability. Users can port identical workflows across computing systems by simply adjusting the configuration files, making rapid prototyping, exploration, and full production campaigns equally straightforward.

\section{Performance and Scalability}
With the use of Parsl, the exa-AMD framework is explicitly designed to leverage modern high-performance computing (HPC) facilities, including exascale or heterogeneous clusters with both CPU and GPU architectures. 
By decoupling workflow logic from execution resources, exa-AMD can efficiently multiplex tens of thousands of independent jobs, automatically scaling workload to available nodes and managing asynchronous dependencies. Importantly, Parsl supports dynamic provisioning and workload elasticity: as the workflow proceeds through different stages in the workflow, computing resources can be grown or shrunk, and jobs can transparently recover or resume from failures without user's intervention. This ensures efficient backfilling and utilization of large, shared supercomputers.

To quantify performance, we conducted systematic scaling tests on three representative systems: Na-B-C, Ce-Co-B, and Fe-Co-Zr. Na-B-C only involves light elements, making it the simplest system among them. Ce-Co-B is relatively complicated, involving rare-earth element as well as 3d transition metal. Fe-Co-Zr is selected as a typical rare-earth-free magnetic system. For each system, we measured total wall-clock time required to complete a workflow consisting of structure generation, CGCNN-based screening, structure selection and a fixed-sized pool of parallelized DFT relaxations. Structure generation and selection were performed on CPUs only, while CGCNN screening and DFT calculations made use of both CPU and GPU architectures depending on the benchmark configuration. Benchmarks were performed on both CPU and GPU partitions, with node counts up to 32 for the Na-B-C system, and up to 256 for the Ce-Co-B and Fe-Co-Zr systems. The significantly smaller node count for Na-B-C reflects the lower computational demands of this system: it involves only light elements (Na, B, C) with fewer electrons and simpler electronic structures, resulting in faster DFT convergence and shorter computation time per structure. In contrast, Ce-Co-B and Fe-Co-Zr are substantially more computationally expensive---Ce-Co-B contains rare-earth element Ce with partially filled 4f orbitals requiring careful electronic treatment, while Fe-Co-Zr involves 3d transition metals with complex magnetic interactions. These systems require more self-consistent-field iterations for convergence, necessitating larger node counts to achieve comparable wall-clock times for benchmarking purposes. 

\begin{figure}[h!tb]
    \centering
    \includegraphics[width=0.7\textwidth]{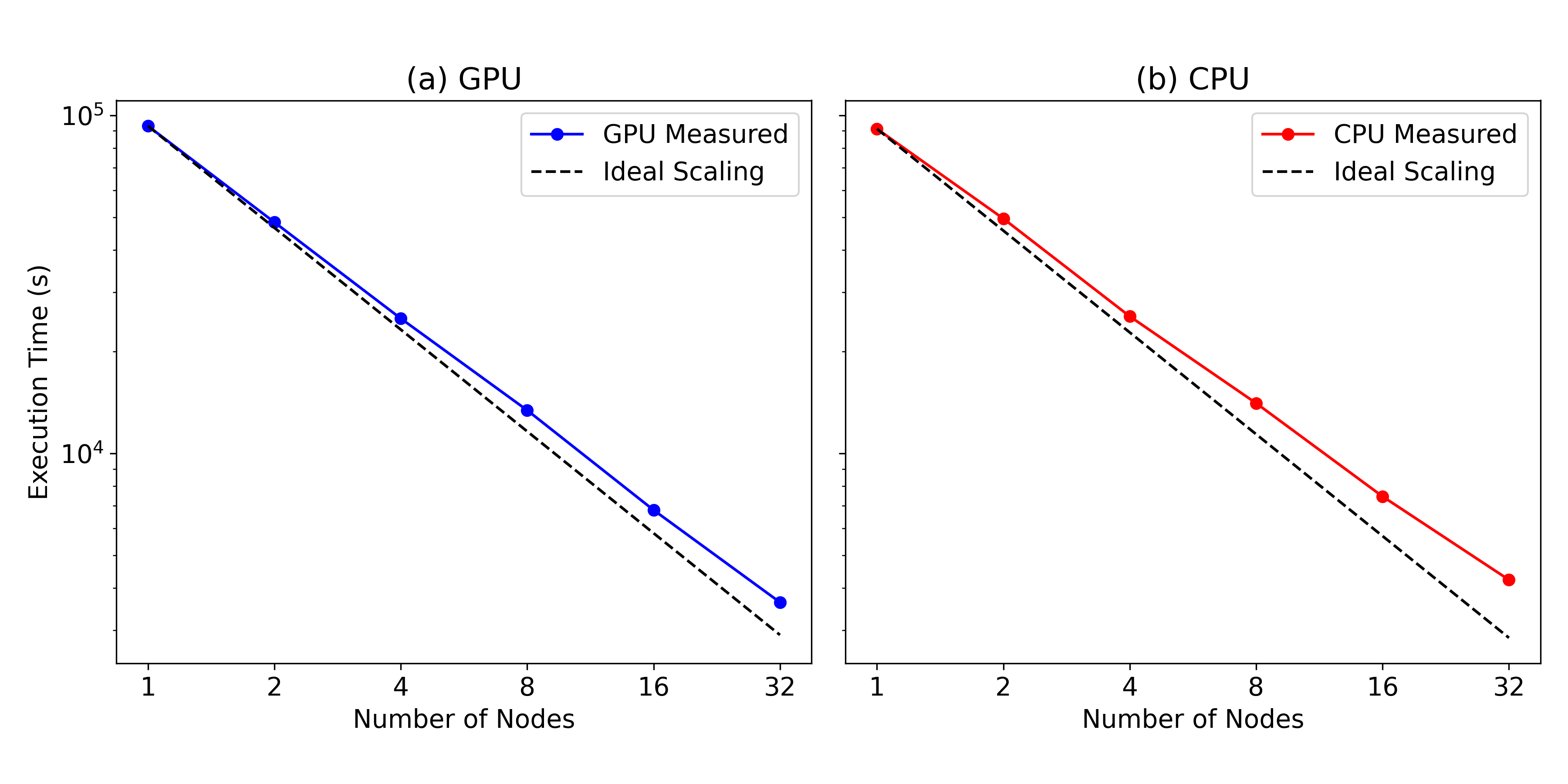}
    \caption{Strong scaling of exa-AMD workflow for the Na-B-C system on NERSC's Perlmutter supercomputer.  Wall-clock times are shown for both GPU and CPU nodes (1, 2, 4, 8, 16, 32). The workflow exhibits near-linear speed-up and substantial GPU acceleration at all node counts.}
    \label{fig:NaBC-benchmark}
\end{figure}

\begin{figure}[h!tb]
    \centering
    \includegraphics[width=0.7\textwidth]{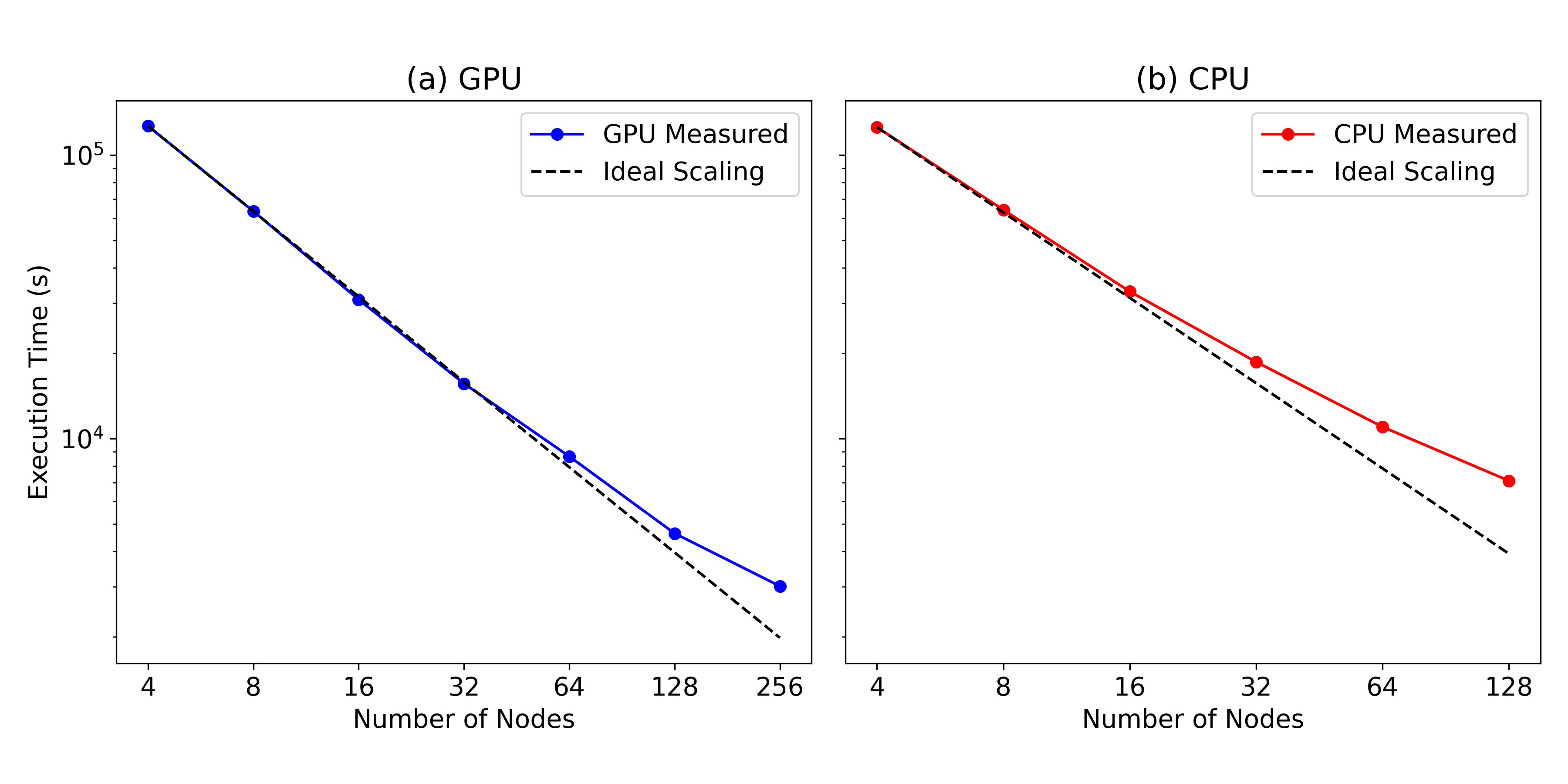}
    \caption{Benchmark results for the Ce-Co-B system on NERSC's Perlmutter supercomputer, executed on GPU architectures for 4 to 256 nodes, and on CPU for 4 to 128 nodes. Performance demonstrates efficient strong scaling and consistent GPU advantage for large-scale campaigns.}
    \label{fig:CeCoB-benchmark}
\end{figure}

\begin{figure}[h!tb]
    \centering
    \includegraphics[width=0.7\textwidth]{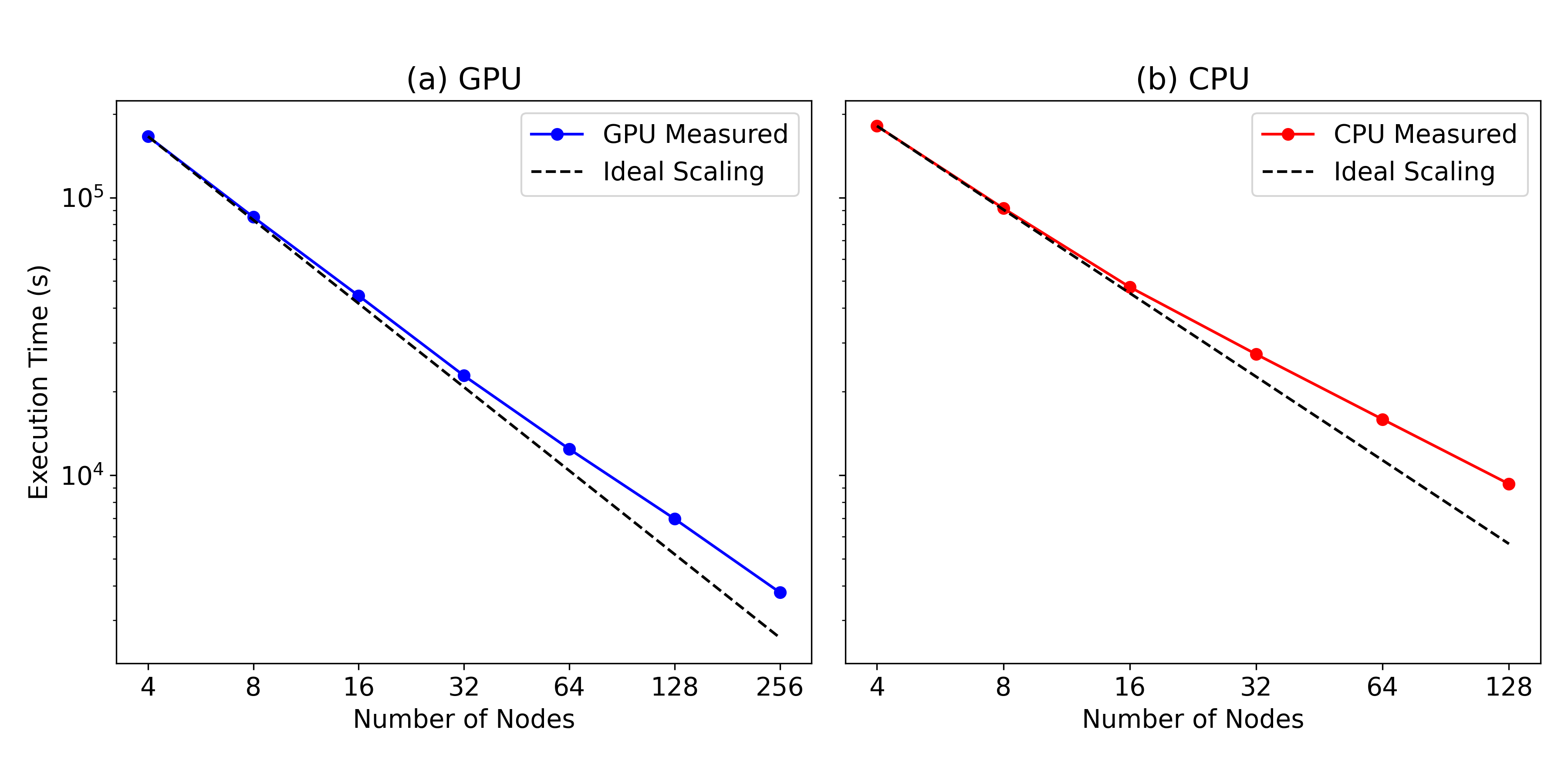}
    \caption{Wall-clock time as a function of node count for the Fe-Co-Zr system on NERSC's Perlmutter supercomputer, on GPU architectures for 4 to 256 nodes, and on CPU for 4 to 128 nodes. Strong scaling is maintained across the full range, underscoring exa-AMD's readiness for large, exascale discovery campaigns.}
    \label{fig:FeCoZr-benchmark}
\end{figure}

Figs.~\ref{fig:NaBC-benchmark}, \ref{fig:CeCoB-benchmark}, and \ref{fig:FeCoZr-benchmark} summarize the strong scaling results: wall-clock times decrease nearly ideally with node count, following a $1/N$ scaling, on both CPU and GPU backends, illustrating excellent parallel efficiency. We performed the benchmark tests on two large computers, Perlmutter at the National Energy Research Scientific Computing Center (NERSC) and Chicoma at Los Alamos National Laboratory (LANL). Each Perlmutter's CPU node is comprised of two AMD EPYC 7763 (Milan) 64-core 2.45 GHz CPUs; each GPU node has one CPU of the same architecture and four NVIDIA A100 GPUs. Chicoma has similar hardware specifications: each CPU node has two AMD EPYC 7H12 (Rome) 64-core 2.6 GHz CPUs; whereas each GPU node has one AMD EPYC 7713 (Rome) 64-core 2 GHz CPU and four NVIDIA A100 GPUs. For example, in the Na-B-C system on NERSC's Perlmutter supercomputer, wall-clock time reduced from 1550 minutes on a single node with 4 GPUs to 88 minutes on 32 GPU nodes, and from 1520 to 98 minutes over the same number of CPU nodes. GPU benchmarks consistently achieve slightly faster runtime than on CPUs.  Moreover, compared to CPU benchmarks, GPU benchmarks exhibit strong scaling that is closer to ideal behavior at higher node counts, reflecting the efficiency of accelerating ML inference and ab initio calculations by modern GPU architectures.

Performance comparisons for the Ce-Co-B and Fe-Co-Zr systems on NERSC's Perlmutter supercomputer reveal similar strong scaling: for Ce-Co-B, total workflow time dropped from 2112 to 50 minutes on GPUs (4 to 256 nodes), and 2091 to 118 minutes on CPUs (4 to 128 nodes). For Ce-Co-B and Fe-Co-Zr benchmarks, we start from 4 nodes rather than the conventional single-node baseline due to practical time constraints. A single-node run would require approximately 2.5–3 days to complete the full workflow for these computationally demanding systems. Such extended runtime are impractical for systematic benchmarking studies and exceed typical time limits on shared HPC resources. By using 4 nodes as our baseline, we maintain reasonable benchmark completion times while still demonstrating strong scaling behavior across two orders of magnitude in node count. The near-ideal $1/N$ scaling observed from 4 to 256 nodes strongly suggests that this scaling behavior would extend to lower node counts, though at the cost of prohibitively long wall-clock times for practical use.
The parallel efficiency typically remains at above 80\% across the tested range. 
The highly modular design managed by Parsl ensures that each job (whether a ML inference, structure relaxation, or post-processing stage) is distributed, tracked, and aggregated without significant idle time or manual intervention. Dynamic resource allocation and robust fault tolerance further maximize throughput and utilization, particularly in shared supercomputing environments with variable queueing latency and hardware availability.

These results directly demonstrate that exa-AMD achieves almost optimal scaling and efficient throughput across the CPU and GPU machines we tested on. This enables million-structure ML screening in under an hour and ab initio relaxation of thousands of candidates within a day on modern clusters. 

\begin{figure}[h!tb]
  \centering

  \includegraphics[width=0.75\textwidth]{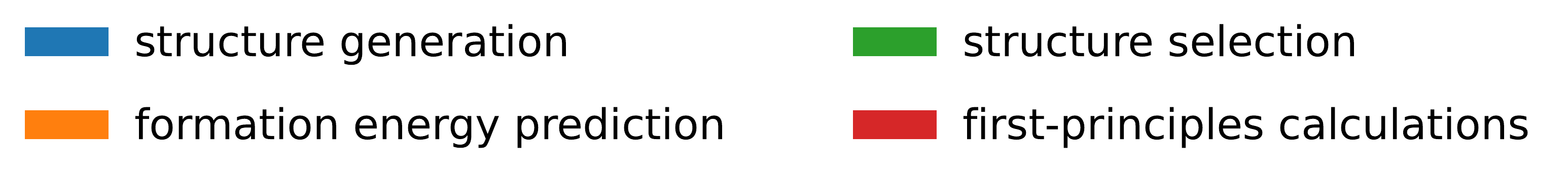}
  \vspace{5pt}

  \begin{subfigure}[t]{0.48\textwidth}
    \centering
    \includegraphics[width=\linewidth]{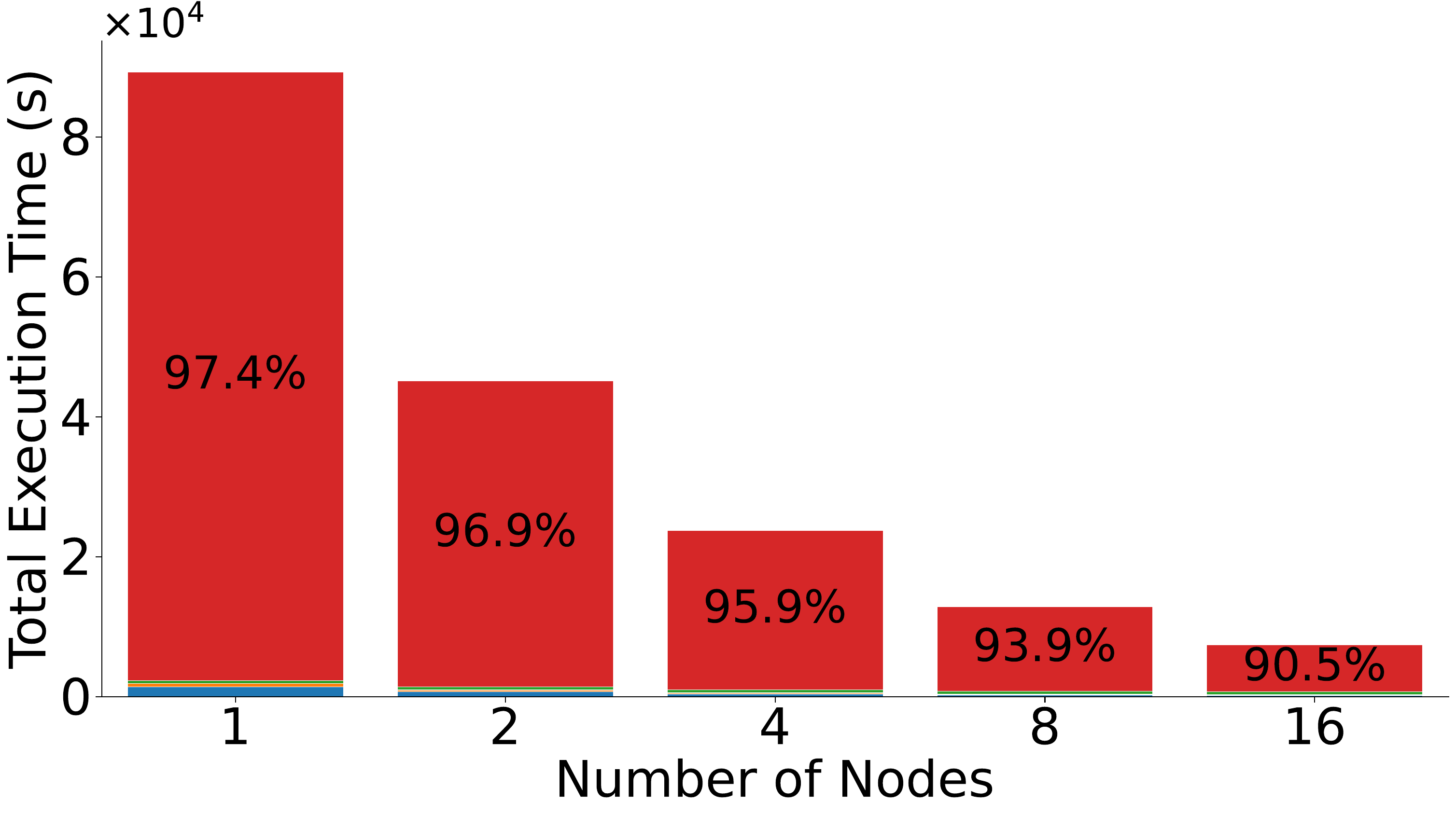}
    \caption{Proportion of time spent on DFT in an end-to-end workflow.}
  \end{subfigure}\hfill
  \begin{subfigure}[t]{0.48\textwidth}
    \centering
    \includegraphics[width=\linewidth]{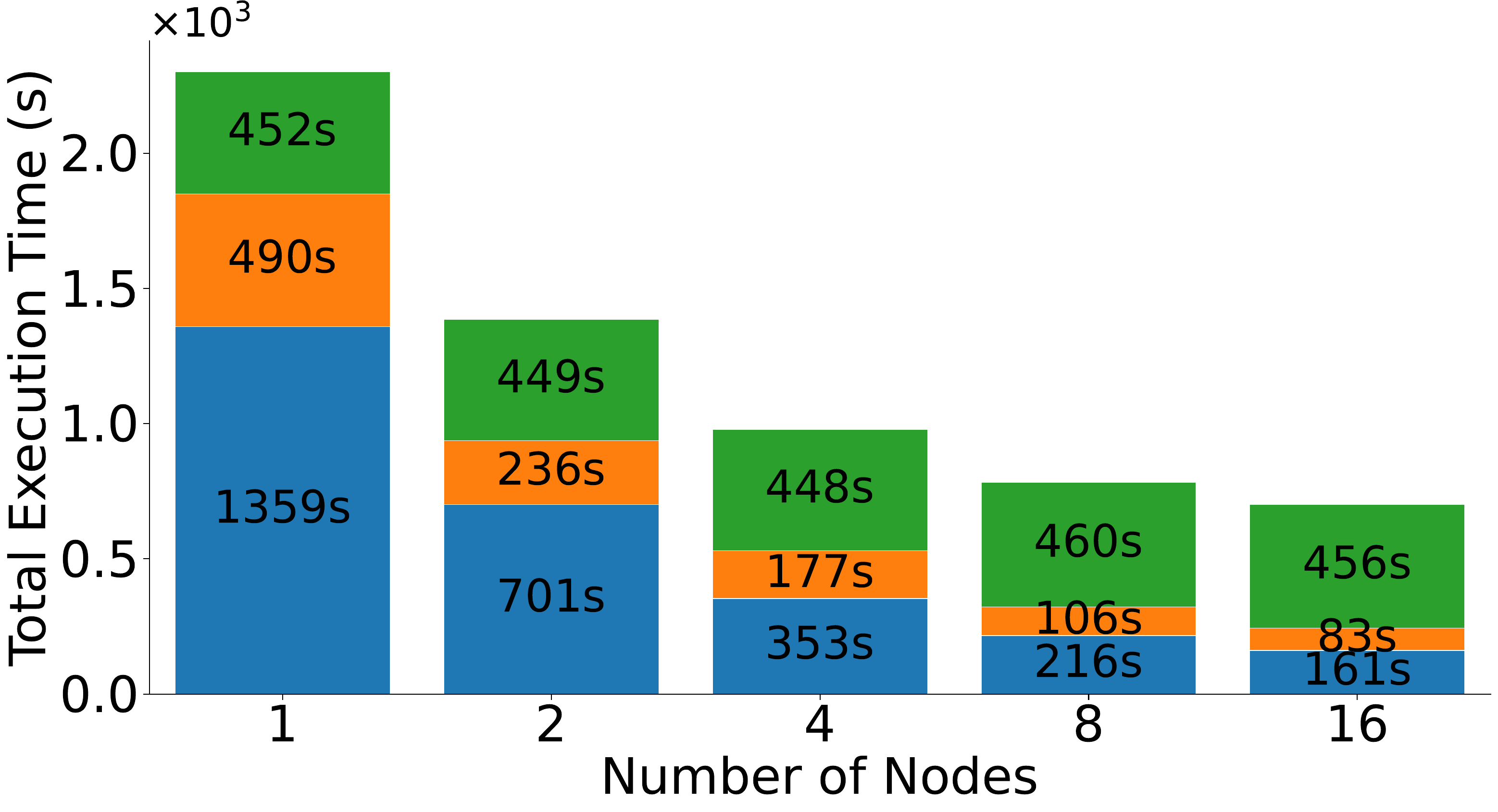}
    \caption{Pre-DFT timing breakdown (absolute time).}
  \end{subfigure}

  \vspace{2pt}
  \caption{Time distribution among major workflow phases for the Na-B-C benchmark performed on LANL's Chicoma supercomputer. Shown are the fractions for structure generation (blue), CGCNN inference (orange), structure selection (green), and VASP DFT calculations (red). For clarity, we show the exact time for pre-DFT calculations, and only the relative percentage (compared to total execution time) for DFT calculations.}
  \label{fig:NaBC-breakdown}
\end{figure}

To examine the time distribution of different phases of the exaAMD workflow, we analyzed the Na-B-C system as a representative case (Fig.~\ref{fig:NaBC-breakdown}). First principles calculations performed with VASP dominate the total execution time, representing about 90.5\% on 16 nodes. The remaining time is divided among structure generation, CGCNN inference for formation energy prediction, and structure selection. The structure selection phase is parallelized only with shared memory and therefore runs on a single node, which limits its scalability relative to the other stages. Nevertheless, even at large scales this phase remains a minor contributor to the total execution time, confirming that the overall cost is governed primarily by the ab initio calculations.
With GPU acceleration, the CGCNN model can process over 1 million structures within minutes, allowing exa-AMD to rapidly down-select candidates and prioritize compute resources for only the most promising compounds. This efficient scalability---made possible by Parsl’s fine-grained, parallel task scheduling---demonstrates that the workflow overhead is minimal and that exa-AMD achieves near-optimal throughput for production-scale discovery, limited primarily by the computationally intensive quantum mechanical calculations.

As the workflow scales to larger node counts, the relative overhead from non-DFT tasks (structure generation, ML inference, and job management) increases modestly due to communication and coordination costs, reaching approximately 15\% at 32 nodes for the Na-B-C system. However, this scaling behavior remains acceptable for production use: the absolute time spent on these tasks decreases with parallelization, and the overall workflow still exhibits strong scaling efficiency (>80\%) across the tested range. By leveraging Parsl's asynchronous task management and dynamic resource allocation, exa-AMD minimizes idle compute time and ensures that the bulk of computational resources remain focused on the high-value DFT calculations throughout the discovery campaign. Resources are automatically released from structure screening phases and reallocated for the surge in DFT job submissions, allowing seamless scaling and minimized overall turnaround time for large discovery efforts.

\section{Example Applications}
\label{example_applications}
A compelling demonstration of the exa-AMD framework is its application to accelerated discovery and design of rare-earth-free permanent magnets in the ternary Fe-Co-Zr system \cite{FeCoZr}. One of the hallmarks of exa-AMD is its user-focused interface and workflow: the user simply specifies a set of chemical elements---in this case, Fe, Co, and Zr---and the framework automates the entire process of generating, screening, and validating hypothetical compounds using the prototype structures provided by exa-AMD by default, requiring minimal manual intervention.

After specifying the elements, the workflow proceeds by generating a broad pool of candidate structures using elemental substitution on template structures and lattice scaling. CGCNN predicts the formation energies for nearly 900,000 candidate Fe-Co-Zr compositions in under 15 minutes using a single GPU node. Structures with predicted formation energy below a set threshold (e.g., $E_\mathrm{f} < -0.1$~eV/atom) are automatically filtered in. After removing duplicated structures, this yields about 3,100 distinct candidates for ab initio validation.
Subsequent first-principles calculations using VASP, including structural relaxation and total energy calculations for all 3,100 structures, were completed in less than 12 hours on a single GPU node with 4 GPUs. The only user input required throughout this discovery pipeline was the choice of target elements. All stability screening, filtering, relaxation, and post-processing were handled automatically, demonstrating both the simplicity and power of the exa-AMD approach.

\begin{figure}[h!tb]
    \centering

    \begin{subfigure}[t]{0.44\textwidth} 
        \centering
        \includegraphics[width=\linewidth,height=5cm]{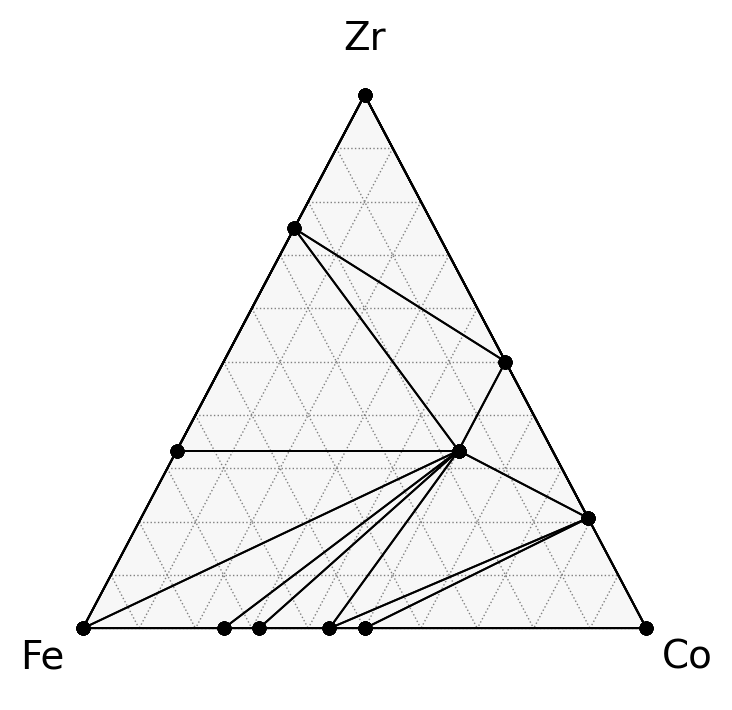} 
        \caption{Known Fe-Co-Zr convex hull.}
        \label{subfig:old-hull}
    \end{subfigure}\hfill
    \begin{subfigure}[t]{0.54\textwidth} 
        \centering
        \raisebox{4pt}{
        \includegraphics[width=\linewidth,height=5cm]{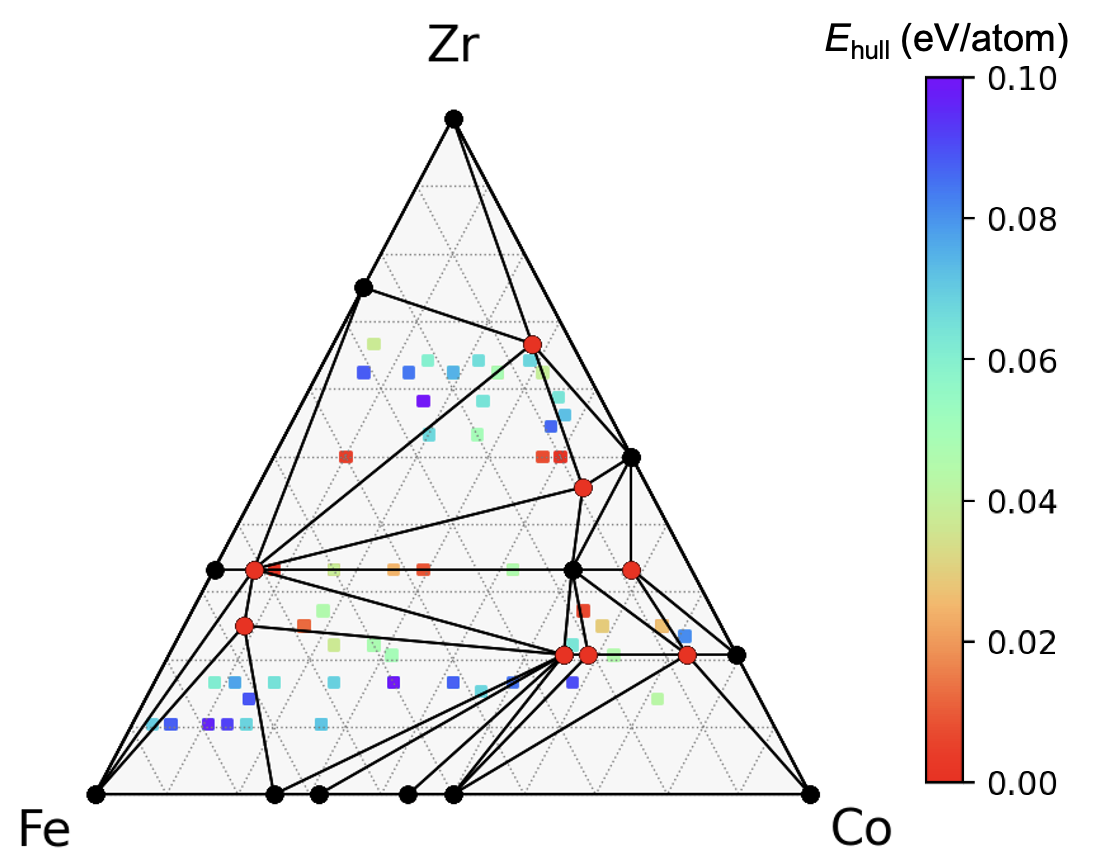}} 
        \caption{Updated Fe-Co-Zr convex hull.}
        \label{subfig:new-hull}
    \end{subfigure}

    \vspace{5pt}
    \caption{
    Comparison of the Fe-Co-Zr convex hull. \textbf{(a)} The convex hull calculated using previously known stable structures from the Materials Project \cite{MP}. \textbf{(b)} The updated convex hull from the exa-AMD framework, showing newly predicted stable and metastable phases. Colored squares in (b) indicate newly identified metastable phases with low formation energies (within 0.1 eV/atom above the convex hull).
}
    \label{fig:FeCoZr-hulls-comparison}
\end{figure}

As a final output, the exa-AMD framework provided a significantly updated convex hull for the Fe-Co-Zr phase diagram  (Fig.~\ref{fig:FeCoZr-hulls-comparison}). The calculated results revealed a dramatically extended known landscape for this key system, including the discovery of nine new stable Fe-Co-Zr phases and 81 promising metastable compounds (those within $0.1$ eV/atom above the hull). Beyond phase stability predictions, exa-AMD provides DFT-optimized crystal structures for these new compounds. Optional postprocessing can be employed for selected candidates to calculate complex materials properties, such as site-specific magnetic properties, electronic band structures, and densities of states when required \cite{FeCoZr}.

\begin{figure}[h!tb]
    \centering
    \begin{subfigure}[t]{0.42\textwidth}
        \centering
        \includegraphics[width=\linewidth,height=5cm]{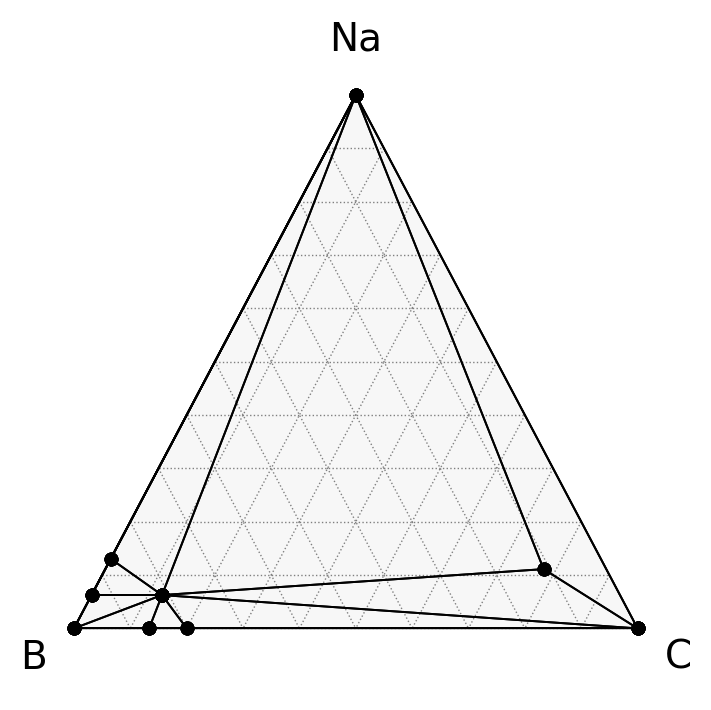} 
        \caption{Known Na-B-C convex hull.}
        \label{subfig:old-NaBC-hull}
    \end{subfigure}\hfill
    \begin{subfigure}[t]{0.56\textwidth}
        \centering
        \raisebox{3pt}{
        \includegraphics[width=\linewidth,height=5cm]{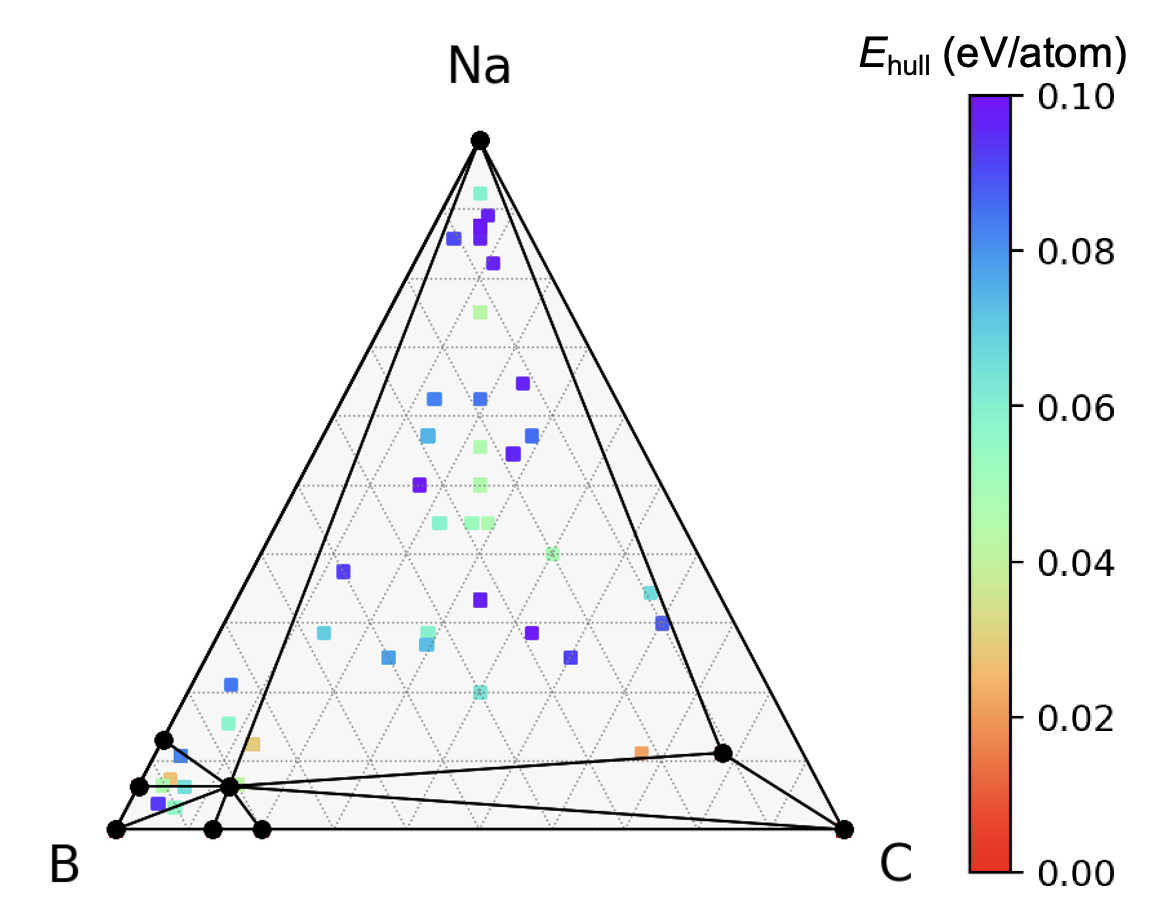}} 
        \caption{Updated Na-B-C convex hull.}
        \label{subfig:new-NaBC-hull}
    \end{subfigure}

    \vspace{5pt}
    \caption{
        Comparison of the Na-B-C convex hull. \textbf{(a)} The previously established phase stability landscape calculated using structures obtained from the Materials Project \cite{MP}. \textbf{(b)} The updated convex hull resulting from the exa-AMD framework. Colored squares in (b) indicate newly identified metastable phases with low formation energies (within 0.1 eV/atom above the convex hull).
    }
    \label{fig:NaBC-hull-comparison}
\end{figure}

The exa-AMD workflow is equally applicable to a wide range of materials and other target properties. For demonstration, the Na-B-C system provides a clear example of the framework's power in a complex chemical space  (Fig.~\ref{fig:NaBC-hull-comparison}). The user again simply specifies the three desired elements. exa-AMD then systematically explores all possibilities generated by template substitution, screens stable candidates using ML models, and refines selected compound energies and properties by DFT calculations. The outputs include both updated convex hulls (for phase stability visualization) and, if desired, relaxed structure information for experimental synthesis or further theoretical study. exa-AMD revised the known thermodynamic landscape for this system, and discovered 11 new metastable Na-B-C phases (within the $0.1$ eV/atom threshold above the hull). The ability to efficiently sample and optimize structures across this ternary space allows us to pinpoint compounds that may possess desirable properties for applications such as energy storage or high-hardness materials.

The discovery of quaternary systems and higher-order multi-principal element alloys (e.g., quinary Al-Co-Cr-Fe-Ni high-entropy alloys) can be pursued using the same high-throughput approach. The framework is particularly valuable for exploring computationally demanding systems that challenge conventional DFT workflows: rare-earth-containing compounds (e.g., Ce-Co-B, Nd-Fe-B, Sm-Co alloys) with strongly correlated 4f electrons requiring careful convergence; actinide-based materials (e.g., U-Mo-Zr nuclear fuels) with 5f orbital complexity and large spin-orbit coupling; systems with heavy transition metals (e.g., Pt-Ir-Rh catalysts, W-Re-Os refractory alloys) demanding dense k-point meshes and high plane-wave cutoffs; and magnetic materials with complex spin configurations (e.g., frustrated magnets, non-collinear magnetic structures in Mn-based alloys). These more complex systems, which involve combinatorially larger composition spaces and substantially increased computational cost per structure, benefit significantly from exa-AMD's automated workflow, GPU acceleration, and efficient scaling to hundreds of nodes, making them particularly suitable for exascale discovery campaigns.
The modular, scalable, and element-agnostic strategy allows exa-AMD users to efficiently perform reliable high-throughput discovery campaigns for virtually any combination of elements, fulfilling a critical need for rapid, reproducible computational materials design.


















\section{User Guide}
This section is not intended to replace the official online documentation of \texttt{exa-AMD}. Instead, it is meant to complement it by providing a concise overview of the installation, configuration, and execution steps, together with practical information that helps new users quickly identify and resolve common issues encountered when setting up or running the workflow.

\subsection{Installation}
\texttt{exa-AMD} is publicly available at \url{https://github.com/ML-AMD/exa-amd}. The package can be installed either
from a prebuilt release or directly from source. The recommended method is to install the distributed wheel:
\begin{lstlisting}
# create & activate a fresh virtual environment
# python -m venv myenv
# source myenv/bin/activate
# python -m pip install --upgrade pip
pip install "https://github.com/ML-AMD/exa-amd/releases/download/v0.1.1/exa_amd-0.1.1-py3-none-any.whl"
exa_amd --help
\end{lstlisting}
For development or testing, one can clone the repository, create the \texttt{Conda} environment defined in
\texttt{amd\_env.yml}, and execute the main module directly:
\begin{lstlisting}
git clone https://github.com/ML-AMD/exa-amd.git
cd exa-amd
conda env create -f amd_env.yml
conda activate amd_env
export PYTHONPATH=$(pwd):$PYTHONPATH
export PYTHONPATH="$(pwd)/ml_models:$PYTHONPATH"
python exa_amd.py --help
\end{lstlisting}
The last command lists all the command line arguments that can be passed into the main executable to control the workflow setting. All software dependencies are specified in \texttt{amd\_env.yml} at the root of the repository and are not repeated here. In particular, a working installation of a first-principles code such as \texttt{VASP} is required for the DFT optimization stages, together with the corresponding VASP input files, e.g., the projector-augmented-wave (PAW) pseudopotentials.

\subsection{Configuration and Execution}
\label{subsec:configuration}

User input is provided through a single JSON configuration file that defines the target chemical system, the relevant working directories, and a reference to a Parsl configuration file that specifies the computational resources. These parameters in the JSON file can be overridden by the command line arguments provided to the main executable \texttt{exa\_amd.py}. For brevity, we do not reproduce the entire schema here since a complete description is available in the online documentation.
Instead, we summarize below the most relevant parameters required for a new user to immediately begin using \texttt{exa-AMD}. As a starting point, users are encouraged to copy and adapt the configuration template \texttt{configs/perlmutter.json} distributed with the repository (this holds even if the target platform is not Perlmutter).

\begin{center}
\begin{tabular}{p{0.29\linewidth}p{0.65\linewidth}}
\texttt{"work\_dir"} & Absolute path to the working directory for the first three stages (structure generation, formation-energy prediction, and structure selection); typically on a scratch filesystem. \\[3pt]

\texttt{"elements"} & Target chemical composition: ternary or quaternary systems (e.g., \texttt{"Na-B-C"} or \texttt{"Na-B-H-C"}). \\[3pt]

\makecell[tl]{\texttt{"initial\_}\\\texttt{structures\_dir"}} & Directory of initial structure templates used to generate hypothetical compounds. A curated set is available at \url{https://zenodo.org/records/17180192}; users may also provide their own. \\[3pt]

\texttt{"parsl\_config"} & Selects a named runtime resource configuration (e.g., \texttt{"perlmutter\_premium"}). \\[3pt]

\texttt{"parsl\_configs\_dir"} & Directory searched at startup for user-authored Parsl configuration files. \\[3pt]

\texttt{"vasp\_std\_exe"} & Executable name or absolute path of the VASP standard binary. \\[3pt]

\texttt{"vasp\_work\_dir"} & Working directory for VASP jobs; separate from \texttt{"work\_dir"} to accommodate systems where scratch has constraints. \\[3pt]

\texttt{"vasp\_pot\_dir"} & Path to the PAW potentials directory for GGA calculations. \\[3pt]

\texttt{"vasp\_output\_file"} & Name of the CSV summarizing structures that successfully converged (e.g., \texttt{"vasp\_results.csv"}). \\[3pt]

\texttt{"vasp\_nstructures"} & Maximum number of structures processed in this run (\texttt{-1} = all). \texttt{exa-AMD} tracks progress and continues from the next structure on subsequent runs (e.g., 1–10, then 11–20). \\[3pt]

\makecell[tl]{\texttt{"post\_processing\_}\\\texttt{output\_dir"}} & Directory for post-processing outputs and summaries; if unset, post-processing is skipped. \\[3pt]

\texttt{"mp\_rester\_api\_key"} & Materials Project API key for reference-data queries during post-processing (required if \texttt{"post\_processing\_output\_dir"} is set). \\
\end{tabular}
\end{center}

\noindent The JSON file also defines general workflow parameters, such as the formation-energy threshold, worker counts, and VASP runtime options, and when using the built-in Parsl configurations it provides helper fields to set Parsl resources (for example, the number of nodes for the DFT calculations stage). 

The Parsl settings above only select a configuration by name and indicate where configurations are discovered. We now outline how these configurations specify resource selection and executor behavior in practice. Indeed, a Parsl configuration describes how different workflow stages are mapped to CPU and GPU resources and how far each stage is allowed to scale. Below we provide a minimal illustrative snippet for a GPU-enabled VASP executor on Perlmutter from the \texttt{PerlmutterConfig} class, which also defines the different executors for other parts of the workflow (CGCNN prediction, structure generation, etc). The full configuration and complete examples are available in the Github repository.
\begin{lstlisting}
class PerlmutterConfig(Config):
    def __init__(self, json_config):
        ...
        # VASP executor
        vasp_executor = HighThroughputExecutor(
            label=VASP_EXECUTOR_LABEL,
            cores_per_worker=1,
            available_accelerators=4,
            provider=SlurmProvider(
                account=gpu_account,
                qos="premium",
                constraint="gpu",
                init_blocks=0,
                min_blocks=0,
                max_blocks=16,
                nodes_per_block=1,
                launcher=SimpleLauncher(),
                walltime='16:00:00',
                worker_init="module load vasp/6.4.3-gpu",
            )
        )
\end{lstlisting}
In this example, we specify \texttt{SlurmProvider} to indicate that the computations will be executed on a Slurm-managed system.
Parsl will create a Slurm script for job submission and requesting for computational resources.
The fields \texttt{account}, \texttt{qos}, and \texttt{constraint} map directly to their \texttt{srun}/\texttt{sbatch} equivalents
on Perlmutter. The setting \texttt{max\_blocks = 16} permits using up to sixteen GPU nodes (subject to the number of
pending tasks), and \texttt{available\_accelerators = 4} schedules one task per GPU, for four concurrent VASP tasks per
node. The label \texttt{VASP\_EXECUTOR\_LABEL} identifies the executor used for the DFT stage of the workflow, meaning
that the relaxation and total-energy calculations are executed using the computational resources defined in this block.
The other executors, used for structure generation, CGCNN inference, selection, and post-processing follow analogous definitions and are documented at \url{https://ml-amd.github.io/exa-amd/workflow.html}.

After defining a full configuration (see the \texttt{parsl\_configs} directory at the root of the \texttt{exa-AMD} repository
for complete examples), it must be registered so that it can be discovered at runtime:
\begin{lstlisting}
register_parsl_config("perlmutter_premium", PerlmutterConfig)
\end{lstlisting}
The file that contains the registrations is then moved into a directory whose path is provided via the JSON parameter
\texttt{"parsl\_configs\_dir"}. This directory may contain multiple files, each with multiple configurations and calls
to \texttt{register\_parsl\_config}; the only requirement is that registration names are unique across all files in the
directory. At startup, \texttt{exa-AMD} reads all files in \texttt{"parsl\_configs\_dir"} and collects the registered
configurations. The user selects one of them at runtime by setting the JSON parameter \texttt{"parsl\_config"} (e.g., \texttt{"perlmutter\_premium"}).

\subsection{Workflow Execution}
Generally, a typical user follows the steps below to run an end-to-end \texttt{exa-AMD} workflow:
\begin{enumerate}
    \item Populate the structure pool with initial templates provided in this work \cite{exaAMD-dataset} or import from a supported materials database.
    \item Edit the JSON input file described in Section \ref{subsec:configuration} specifying target elements and workflow parameters.
    \item Prepare a Parsl configuration for the target computing platform.
    \item Launch the workflow via command line or Python API.
    \item Monitor progress via logs; restart is possible from saved stages for long or batch jobs.
    \item Analyze the results in the output directory: stable/metastable candidates, updated convex hull, relaxed structures, property tables.
\end{enumerate}
\subsection{Examples and Troubleshooting}
\begin{itemize}
    \item Example workflows and datasets are included in the GitHub documentation site and online guides.
    \item Troubleshooting tips for common errors (convergence, resource allocation, file I/O) are detailed in the user documentation. The tutorial provides an example of predicting novel Na–B–H–C compounds (\url{https://ml-amd.github.io/exa-amd/tutorial.html}).
\end{itemize}

\section{Summary and Outlook}
exa-AMD represents a significant advancement in the field of computational materials science by providing an exascale-capable, modular framework that dramatically accelerates the exploration of complex composition-structure-property spaces. By automating structure generation, stability screening using state-of-the-art machine learning methods, and rigorous first-principles (DFT) calculations in a high-throughput manner using Parsl, exa-AMD enables researchers to efficiently identify new stable and metastable compounds from vast candidate pools. Its capability is demonstrated by a few selected, challenging multinary systems, achieving rapid down-selection and efficient resource utilization with minimal manual intervention. One of the goals of exa-AMD is to support transparent, reproducible, and community-driven research by providing open-source codes with detailed documentation, as well as a user-friendly design.

At present, exa-AMD focuses on first principles characterization on carefully selected compounds and phase diagram refinement based on known structural templates. A limitation is that the proposed compounds are restricted to the structure types present in the existing templates or user-supplied prototypes, making the discovery of entirely new structure motifs unfeasible. To address this problem, future developments for exa-AMD will incorporate two new features: (1) the integration of advanced machine learning potentials for efficient atomistic simulation and structure optimization, and (2) the combination of Adaptive Genetic Algorithm for new structure generation~\cite{AGA1,AGA2}. These new features will enable the prediction of novel crystal structures beyond the predefined prototypes, providing a more robust and comprehensive approach for material discovery.


\section*{Acknowledgements}
Work at Ames National Laboratory and Los Alamos National Laboratory was supported by the U.S. Department of Energy (DOE), Office of Science, Basic Energy Sciences, Materials Science and Engineering Division through the Computational Material Science Center program. Ames National Laboratory is operated for the U.S. DOE by Iowa State University under contract No. DE-AC02-07CH11358. Los Alamos National Laboratory is operated by Triad National Security, LLC, for the National Nuclear Security Administration of U.S. Department of Energy under Contract No. 89233218CNA000001. This research used resources of the National Energy Research Scientific Computing Center (NERSC), a DOE Office of Science User Facility supported under Contract No. DE-AC02-05CH11231, and resources provided by the Los Alamos National Laboratory Institutional Computing Program, which is supported by the U.S. Department of Energy National Nuclear Security Administration under Contract No. 89233218CNA000001. (LA-UR-25-28449)

\bibliographystyle{elsarticle-num}
\bibliography{references}

@article{JOSS,
    author = {Moraru, Maxim and Xia, Weiyi and Ye, Zhuo and Zhang, Feng and Yao, Yongxin and Li, Ying Wai and Wang, Cai-Zhuang},
    title = {{exa-AMD: A Scalable Workflow for Accelerating AI-Assisted Materials Discovery and Design}},
    year = {2025},
    journal = {arXiv preprint arXiv:2506.21449},
    eprint = {2506.21449},
    archiveprefix = {arXiv},
    primaryclass = {cs.DC}
}

@article{Parsl,
    author = {Babuji, Yadu and Woodard, Anna and Li, Zhuozhao and Katz, Daniel S. and Clifford, Ben and Kumar, Rohan and Lacinski, Lukasz and Chard, Ryan and Wozniak, Justin M. and Foster, Ian},
    title = {Parsl: Pervasive Parallel Programming in Python},
    year = {2019},
    issue_date = {June 2019},
    publisher = {Association for Computing Machinery},
    address = {New York, NY, USA},
    volume = {28},
    journal = {Proceedings of the 28th International Symposium on High-Performance Parallel and Distributed Computing},
    pages = {25–36},
    doi = {10.1145/3307512.3329361}
}

@article{FeCoZr,
    author = {Xia, Weiyi and Moraru, Maxim and Li, Ying Wai and Liao, Timothy and Chelikowsky, James R. and Wang, Cai-Zhuang},
    title = {{Accelerated discovery and design of Fe-Co-Zr magnets with tunable magnetic anisotropy through machine learning and parallel computing}},
    year = {2025},
    journal = {arXiv preprint arXiv:2506.22627},
    eprint = {2506.22627},
    archiveprefix = {arXiv},
    primaryclass = {cond-mat.mtrl-sci}
}

@article{Sun2022,
  author = {H. Sun and C. Zhang and W. Xia and L. Tang and R. Wang and G. Akopov and N. W. Hewage and K.-M. Ho and K. Kovnir and C.-Z. Wang},
  title = {Machine Learning-Guided Discovery of Ternary Compounds Containing La, P, and Group 14 Elements},
  journal = {Inorganic Chemistry},
  year = {2022},
  volume = {61},
  number = {42},
  pages = {16699--16706},
  doi = {10.1021/acs.inorgchem.2c02431}
}

@article{PNAS,
  author = {W. Xia and M. Sakurai and B. Balasubramanian and T. Liao and R. Wang and C. Zhang and H. Sun and K.-M. Ho and J. R. Chelikowsky and D. J. Sellmyer and C.-Z. Wang},
  title = {Accelerating the discovery of novel magnetic materials using a machine learning guided adaptive feedback},
  journal = {Proceedings of the National Academy of Sciences},
  year = {2022},
  volume = {119},
  number = {47},
  pages = {e2204485119},
  doi = {10.1073/pnas.2204485119}
}

@article{LaCoPb,
  author = {R. Wang and W. Xia and X. Fan and H. Dong and T. J. Slade and K.-M. Ho and P. C. Canfield and C.-Z. Wang},
  title = {ML-guided discovery of La-Co-Pb ternary compounds involving immiscible pairs of Co and Pb elements},
  journal = {npj Computational Materials},
  year = {2023},
  volume = {8},
  number = {1},
  pages = {258},
  doi = {10.1038/s41524-022-00950-0}
}

@article{FeCoSi,
  author = {T. Liao and W. Xia and M. Sakurai and R. Wang and C. Zhang and H. Sun and K.-M. Ho and C.-Z. Wang and J. R. Chelikowsky},
  title = {Magnetic iron-cobalt silicides discovered using machine-learning},
  journal = {Physical Review Materials},
  year = {2023},
  volume = {7},
  number = {3},
  pages = {034410},
  doi = {10.1103/PhysRevMaterials.7.034410}
}

@article{JMCA,
  author = {W. Xia and L. Tang and H. J. Sun and C. Zhang and K.-M. Ho and G. Viswanathan and K. Kovnir and C.-Z. Wang},
  title = {Accelerating materials discovery using integrated deep machine learning approaches},
  journal = {Journal of Materials Chemistry A},
  year = {2023},
  volume = {11},
  number = {47},
  pages = {25973--25982},
  doi = {10.1039/D3TA03771A}
}

@article{FeCoC,
  author = {W. Xia and M. Sakurai and B. Balasubramanian and T. Liao and R. Wang and C. Zhang and H. Sun and K.-M. Ho and J. R. Chelikowsky and C.-Z. Wang},
  title = {Machine learning assisted search for Fe-Co-C ternary compounds with high magnetic anisotropy},
  journal = {APL Machine Learning},
  year = {2024},
  volume = {2},
  number = {4},
  pages = {046103},
  doi = {10.1063/5.0185935}
}

@article{FeCoP,
  author = {T. Liao and W. Xia and M. Sakurai and C. Zhang and H. Sun and R. Wang and K.-M. Ho and C.-Z. Wang and J. R. Chelikowsky},
  title = {Machine learning-accelerated discovery of iron cobalt phosphides as rare-earth-free magnets},
  journal = {Physical Review Materials},
  year = {2024},
  volume = {8},
  pages = {104404},
  doi = {10.1103/PhysRevMaterials.8.104404}
}

@article{CeFeX,
  author = {W. Xia and W. Tee and P. C. Canfield and F. Garcia and R. Ribeiro and Y. Lee and L. Ke and R. Flint and C.-Z. Wang},
  title = {Machine learning accelerated prediction of Ce-based ternary compounds involving antagonistic pairs},
  journal = {Physical Review Materials},
  year = {2025},
  volume = {9},
  number = {5},
  pages = {053803},
  doi = {10.1103/PhysRevMaterials.9.053803}
}

@article{CeCoCu,
  author = {W. Xia and W. S. Tee and P. C. Canfield and R. Flint and C.-Z. Wang},
  title = {Search for stable and metastable Ce-Co-Cu ternary compounds using machine learning},
  journal = {Inorganic Chemistry},
  year = {2025},
  note = {Just published, DOI not yet assigned}
}

@article{MP,
  author = {A. Jain and S. P. Ong and G. Hautier and W. Chen and W. D. Richards and S. Dacek and S. Cholia and D. Gunter and D. Skinner and G. Ceder and K. A. Persson},
  title = {Commentary: The Materials Project: A materials genome approach to accelerating materials innovation},
  journal = {APL Materials},
  year = {2013},
  volume = {1},
  pages = {011002},
  doi = {10.1063/1.4812323}
}

@article{GNOME,
  author = {A. Merchant and S. Batzner and S. S. Schoenholz},
  title = {Scaling deep learning for materials discovery},
  journal = {Nature},
  year = {2023},
  volume = {624},
  pages = {80--85},
  doi = {10.1038/s41586-023-06735-9}
}

@article{AFLOW,
  author = {S. Curtarolo and W. Setyawan and G. L. W. Hart and M. Jahnatek and R. V. Chepulskii and R. H. Taylor and S. Wang and J. Xue and K. Yang and O. Levy and M. J. Mehl and H. T. Stokes and D. O. Demchenko and D. Morgan},
  title = {AFLOW: An automatic framework for high-throughput materials discovery},
  journal = {Computational Materials Science},
  year = {2012},
  volume = {58},
  pages = {218--226},
  doi = {10.1016/j.commatsci.2012.02.005}
}

@article{OQMD,
  author = {J. E. Saal and S. Kirklin and M. Aykol and B. Meredig and C. Wolverton},
  title = {Materials Design and Discovery with High-Throughput Density Functional Theory: The Open Quantum Materials Database (OQMD)},
  journal = {JOM},
  year = {2013},
  volume = {65},
  pages = {1501--1509},
  doi = {10.1007/s11837-013-0755-4}
}

@article{OQMD2,
  author = {S. Kirklin and J. E. Saal and B. Meredig and A. Thompson and J. W. Doak and M. Aykol and S. Rühl and C. Wolverton},
  title = {The Open Quantum Materials Database (OQMD): assessing the accuracy of DFT formation energies},
  journal = {npj Computational Materials},
  year = {2015},
  volume = {1},
  pages = {15010},
  doi = {10.1038/npjcompumats.2015.10}
}

@article{CGCNN,
  author = {T. Xie and J. C. Grossman},
  title = {Crystal Graph Convolutional Neural Networks for an Accurate and Interpretable Prediction of Material Properties},
  journal = {Physical Review Letters},
  year = {2018},
  volume = {120},
  pages = {145301},
  doi = {10.1103/PhysRevLett.120.145301}
}

@article{ALIGNN,
  title={ALIGNN: an atomistic line graph neural network for improved materials property predictions},
  author={Choudhary, Kamal and DeCost, Brian},
  journal={npj Computational Materials},
  volume={7},
  number={1},
  pages={1--8},
  year={2021},
  publisher={Nature Publishing Group},
  doi={10.1038/s41524-021-00650-1}
}

@article{M3GNET,
  title={A universal graph deep learning interatomic potential for the periodic table},
  author={Chen, Chi and Ong, Shyue Ping},
  journal={Nature Computational Science},
  volume={2},
  number={11},
  pages={718--728},
  year={2022},
  publisher={Nature Publishing Group},
  doi={10.1038/s43588-022-00349-3}
}

@article{MEGNET,
  title={Graph networks as a universal machine learning framework for molecules and crystals},
  author={Chen, Chi and Ye, Weike and Zuo, Yunxing and Zheng, Chen and Ong, Shyue Ping},
  journal={Chemistry of Materials},
  volume={31},
  number={9},
  pages={3564--3572},
  year={2019},
  publisher={ACS Publications},
  doi={10.1021/acs.chemmater.9b01294}
}

@article{CHGNET,
  title={CHGNet: A graph neural network with directional message passing and atomic-cluster-based local descriptors for the condensed phase},
  author={Deng, Bosen and Zhong, Peichen and Jun, Heo and Chen, Chi and Ong, Shyue Ping},
  journal={Nature Machine Intelligence},
  volume={5},
  number={7},
  pages={792--801},
  year={2023},
  publisher={Nature Publishing Group},
  doi={10.1038/s42256-023-00682-2}
}

@article{pymatgen,
  title={Python Materials Genomics (pymatgen): A robust, open-source python library for materials analysis},
  author={Ong, Shyue Ping and Richards, William Davidson and Jain, Anubhav and Hautier, Geoffroy and Kocher, Michael and Cholia, Shreyas and Gunter, Dan and Chevrier, Vincent L and Persson, Kristin A and Ceder, Gerbrand},
  journal={Computational Materials Science},
  volume={68},
  pages={314--319},
  year={2013},
  publisher={Elsevier},
  doi={10.1016/j.commatsci.2012.10.028}
}

@article{convexhull,
  title={Li- Fe- P- O2 phase diagram from first principles calculations},
  author={Ong, Shyue Ping and Wang, Lei and Kang, Byoungwoo and Ceder, Gerbrand},
  journal={Chemistry of Materials},
  volume={20},
  number={5},
  pages={1798--1807},
  year={2008},
  publisher={ACS Publications}
}

@article{VASPa,
  author = {G. Kresse and J. Furthmüller},
  title = {Efficiency of ab-initio total energy calculations for metals and semiconductors using a plane-wave basis set},
  journal = {Computational Materials Science},
  year = {1996},
  volume = {6},
  pages = {15--50},
  doi = {10.1016/0927-0256(96)00008-0}
}

@article{VASPb,
  author = {G. Kresse and J. Furthmüller},
  title = {Efficient iterative schemes for ab initio total-energy calculations using a plane-wave basis set},
  journal = {Physical Review B},
  year = {1996},
  volume = {54},
  pages = {11169--11186},
  doi = {10.1103/PhysRevB.54.11169}
}

@article{PAW,
    author = {Blöchl, P. E.},
    title = {Projector augmented-wave method},
    journal = {Physical Review B},
    volume = {50},
    issue = {24},
    pages = {17953--17979},
    year = {1994},
    doi = {10.1103/PhysRevB.50.17953}
}

@article{GGA,
    author = {Perdew, John P. and Burke, Kieron and Ernzerhof, Matthias},
    title = {Generalized Gradient Approximation Made Simple},
    journal = {Physical Review Letters},
    volume = {77},
    issue = {18},
    pages = {3865--3868},
    year = {1996},
    doi = {10.1103/PhysRevLett.77.3865}
}

@article{QEa,
  author = {P. Giannozzi and O. Andreussi and T. Brumme and O. Bunau and M. Buongiorno Nardelli and M. Calandra and R. Car and C. Cavazzoni and D. Ceresoli and M. Cococcioni and N. Colonna and I. Carnimeo and A. Dal Corso and S. de Gironcoli and P. Delugas and R. A. DiStasio Jr and A. Ferretti and A. Floris and G. Fratesi and G. Fugallo and R. Gebauer and U. Gerstmann and F. Giustino and T. Gorni and J. Jia and M. Kawamura and H.-Y. Ko and A. Kokalj and E. Küçükbenli and M. Lazzeri and M. Marsili and N. Marzari and F. Mauri and N. L. Nguyen and H.-V. Nguyen and A. Otero-de-la-Roza and L. Paulatto and S. Poncé and D. Rocca and R. Sabatini and B. Santra and M. Schlipf and A. P. Seitsonen and A. Smogunov and I. Timrov and T. Thonhauser and P. Umari and N. Vast and X. Wu and S. Baroni},
  title = {Advanced capabilities for materials modelling with Quantum ESPRESSO},
  journal = {Journal of Physics: Condensed Matter},
  year = {2017},
  volume = {29},
  pages = {465901},
  doi = {10.1088/1361-648X/aa8f79}
}

@article{QEb,
  author = {P. Giannozzi and S. Baroni and N. Bonini and M. Calandra and R. Car and C. Cavazzoni and D. Ceresoli and G. L. Chiarotti and M. Cococcioni and I. Dabo and A. Dal Corso and S. Fabris and G. Fratesi and S. de Gironcoli and R. Gebauer and U. Gerstmann and C. Gougoussis and A. Kokalj and M. Lazzeri and L. Martin-Samos and N. Marzari and F. Mauri and R. Mazzarello and S. Paolini and A. Pasquarello and L. Paulatto and C. Sbraccia and S. Scandolo and G. Sclauzero and A. P. Seitsonen and A. Smogunov and P. Umari and R. M. Wentzcovitch},
  title = {QUANTUM ESPRESSO: a modular and open-source software project for quantum simulations of materials},
  journal = {Journal of Physics: Condensed Matter},
  year = {2009},
  volume = {21},
  pages = {395502},
  doi = {10.1088/0953-8984/21/39/395502}
}

@article{AGA1,
    author = {Wu, S. Q. and Ji, M. and Wang, C.-Z. and Nguyen, M. C. and Zhao, X. and Umemoto, K. and Wentzcovitch, R. M. and Ho, K.-M.},
    title = {{An adaptive genetic algorithm for crystal structure prediction}},
    journal = {Journal of Physics: Condensed Matter},
    volume = {26},
    number = {3},
    pages = {035402},
    year = {2014},
    doi = {10.1088/0953-8984/26/3/035402}
}

@article{AGA2,
    author = {Zhao, X. and Nguyen, M. C. and Zhang, W. Y. and Wang, C. Z. and Kramer, M. J. and Sellmyer, D. J. and Li, X. Z. and Zhang, F. and Ke, L. Q. and Antropov, V. P. and Ho, K. M.},
    title = {{Exploring the Structural Complexity of Intermetallic Compounds by an Adaptive Genetic Algorithm}},
    journal = {Physical Review Letters},
    volume = {112},
    issue = {4},
    pages = {045502},
    year = {2014},
    doi = {10.1103/PhysRevLett.112.045502}
}

@article{Novomag,
  author    = {Sakurai, M. and Wang, R. and Liao, T. and Zhang, C. and Sun, H. and Sun, Y. and Wang, H. and Zhao, Xin and Wang, S. and Balasubramanian, B. and Xu, X. and Sellmyer, D. J. and Antropov, V. and Zhang, J. and Wang, C.-Z. and Ho, K.-M. and Chelikowsky, J. R.},
  title     = {Discovering rare-earth-free magnetic materials through the development of a database},
  journal   = {Physical Review Materials},
  volume    = {4},
  issue     = {11},
  pages     = {114408},
  year      = {2020},
  month     = {Nov},
  publisher = {American Physical Society},
  doi       = {10.1103/PhysRevMaterials.4.114408}
}

@article{ML-materials-review,
  author  = {Butler, Keith T. and Davies, Daniel W. and Cartwright, Hugh and Isayev, Olexandr and Walsh, Aron},
  title   = {Machine learning for molecular and materials science},
  journal = {Nature},
  volume  = {559},
  number  = {7715},
  pages   = {547--555},
  year    = {2018},
  doi     = {10.1038/s41586-018-0337-2}
}

@article{high-throughput-review1,
  author  = {Curtarolo, S. and Hart, G. L. W. and Nardelli, M. B. and Mingo, N. and Sanvito, S. and Levy, O.},
  title   = {The high-throughput highway to computational materials design},
  journal = {Nature Materials},
  volume  = {12},
  number  = {3},
  pages   = {191--201},
  year    = {2013},
  doi     = {10.1038/nmat3568}
}

@article{high-throughput-review2,
title = {Finding the needle in the haystack: Materials discovery and design through computational ab initio high-throughput screening},
journal = {Computational Materials Science},
volume = {163},
pages = {108-116},
year = {2019},
issn = {0927-0256},
doi = {https://doi.org/10.1016/j.commatsci.2019.02.040},
url = {https://www.sciencedirect.com/science/article/pii/S0927025619301156},
author = {Geoffroy Hautier},
}

@dataset{exaAMD-dataset,
  author       = {Xia, Weiyi and Wang, Cai-Zhuang},
  title        = {Prototype crystal structures for exa-AMD framework},
  year         = {2025},
  publisher    = {Zenodo},
  doi          = {10.5281/zenodo.17180192},
  url          = {https://doi.org/10.5281/zenodo.17180192},
  version      = {0.1}
}

\end{document}